\documentclass[english]{article}
\usepackage[T1]{fontenc}
\usepackage[utf8]{inputenc}
\usepackage{geometry}
\usepackage{color}
\usepackage{babel}
\usepackage{float}
\usepackage{amsmath}
\usepackage{amsthm}
\usepackage{graphicx}
\usepackage{setspace}
\usepackage[unicode=true,pdfusetitle,
 bookmarks=true,bookmarksnumbered=false,bookmarksopen=false,
 breaklinks=false,pdfborder={0 0 1},backref=false,colorlinks=false]
 {hyperref}

\makeatletter
\theoremstyle{plain}
\newtheorem{lyxalgorithm}{\protect\algorithmname}

\addto\captionsenglish{%
}

\usepackage{fancyhdr}
\makeatother

\providecommand{\algorithmname}{Algorithm}

\begin{document}
\begin{doublespace}
\begin{center}
\textbf{\Large{}Arguably Adequate Aqueduct Algorithm: Crossing A Bridge-Less
Block-Chain Chasm}{\Large\par}
\par\end{center}

\begin{center}
\textbf{\large{}Ravi Kashyap (ravi.kashyap@stern.nyu.edu)}\textbf{ }
\par\end{center}

\begin{center}
\textbf{\large{}Estonian Business School / City University of Hong
Kong}{\large\par}
\par\end{center}

\begin{center}
\begin{center}
\today
\par\end{center}
\par\end{center}

\begin{center}
\textbf{\textcolor{blue}{\href{https://doi.org/10.1016/j.frl.2023.104421}{Edited Version: Kashyap, R. (2023).  Arguably Adequate Aqueduct Algorithm: Crossing A Bridge-Less Block-Chain Chasm.  Finance Research Letters,  September 2023,  104421. }}}
\par\end{center}

\begin{center}
\par\end{center}

\begin{center}
Journal of Economic Literature Codes: D7: Analysis of Collective Decision-Making;
D8: Information, Knowledge, and Uncertainty; I31: General Welfare,
Well-Being; O3 Innovation ,Research and Development, Technological
Change, Intellectual Property Rights
\par\end{center}

\begin{center}
studies
\par\end{center}

\begin{center}
Association for Computing Machinery Classification System: C.2.4:
Distributed Systems; I.2.8: Problem Solving; D.2.11: Software Architectures;
J.4: Computer Applications;K.6.5: Security and Protection; K.4.2:
Social Issues
\par\end{center}

\begin{center}
\pagebreak{}
\par\end{center}

\begin{center}
\tableofcontents{}\pagebreak{}
\par\end{center}
\end{doublespace}

\section{\label{sec:Abstract}Abstract}

We consider the problem of being a cross-chain wealth management platform
with deposits, redemptions and investment assets across multiple networks.
We discuss the need for blockchain bridges to facilitates fund flows
across platforms. We point out several issues with existing bridges.
We develop an algorithm - tailored to overcome current constraints
- that dynamically changes the utilization of bridge capacities and
hence the amounts to be transferred across networks. We illustrate
several scenarios using numerical simulations.

\section{\label{sec:Building-Bridges-That}Building Bridges That Do Not Burn}

With the development of several blockchain platforms, investors will
seek diversified returns to mitigate the risks from investing in one
particular network (Lindman et al., 2017; Kuo et al., 2019; Lu 2019;
Prewett et al., 2020; Zamani et al., 2020; Briola et al., 2023; End-note
\ref{enu:Many-protocols-with}). Service providers will focus on rolling
out various products on different chains - which become investment
opportunities. The complexity of managing funds - and risks - across
multiple platforms will give rise to specialized blockchain wealth
managers similar to mutual funds and hedge funds in traditional finance
(Cai 2018; Peterson 2018; Arshadi 2019; Schär 2021; Kashyap 2021-I;
2021-II; Dos Santos et al., 2022; Agarwal et al., 2009; Stulz 2007;
End-note \ref{enu:Excessive-financialization--}). Investment vehicles
mushrooming on different chains, will require the ability to transfers
funds from one network to another. 

To elaborate further, blockchain-funds will invest in several assets
across different networks. The fund price - to deposit and redeem
wealth - will be the same across all the networks on which the investment
infrastructure will be deployed. Two factors determine the fund price:
1) the combined total value locked (TVL: End-note \ref{enu:TVL})
across all networks, and 2) the number of tokens issued for that fund
across all networks. For example, an investor depositing \$50,000
USD on only one network will be getting exposure to the performance
- returns and diversification benefits - of all assets held across
all networks in that fund. To continually monitor such a portfolio
spread across networks, and change it based on market conditions,
would be an extremely arduous task - almost impossible for non-sophisticated
investors. Kashyap (2022) provides detailed examples of blockchain
investment problems and how many best practices - fund and risk management
- from traditional finance can be adapted for the blockchain realm.

From a network exposure point of view, the entire amount of funds
under management will be seen from two perspectives: 1) network portfolio
- assets on only one platform, and 2) global portfolio that aggregates
all the network portfolios. We need to monitor the weights of assets
globally and strict risk management limits have to apply to the global
portfolio. This global capacity on each asset will be filled by positions
on each network depending on how easily funds can flow between networks
via blockchain bridges (Belchior et al., 2021; Qasse et al., 2019;
Schulte et al., 2019; Hafid et al., 2020; Zhou et al., 2020; Stone
2021; Li et al., 2022; End-note \ref{enu:Blockchain-Bridges}). The
amount of funds transferred across networks will depend on: bridge
capacity, relative network gas fees, trading liquidity, investment
fund flows, asset availability and asset exposure on each network
(Bender et al., 2010; Bass et al., 2017; Liu 2019; Monrat et al.,
2019; Zarir et al., 2021; Bertsimas \& Lo 1998; Almgren \& Chriss
2001; Fung et al., 2022).

Section (\ref{subsec:Bridge-Background-and}) explains the high-level
idea behind properly utilizing available bridge capacities. Section
(\ref{sec:Transferring-Assets-Across-Networks-Algorithm}) provides
the detailed bridge algorithm. Section (\ref{sec:Conclusion}) lists
areas for improvement and concludes. Section (\ref{sec:Numerical-Illustrations})
has numerical illustrations of several scenarios that might occur
in practice. Notes and supplementary explanations have also been provided
in Section (\ref{sec:Explanations-and-End-notes}) for the concepts
mentioned in the main text.

\section{\label{subsec:Bridge-Background-and}Blockchain Bridge Background
Basics}

\textbf{\textit{At present, blockchain bridges act as both a bottleneck
and an Achilles heel.}}

We consider the problem of being a cross chain asset management platform
when there are limitations on the amount of funds - and types of assets
- that can be transferred via bridges. More and bigger bridges will
be built - as time passes - easing bridge capacity issues once the
traffic on the bridges will increase. But right now, there are strict
limits on how much funds can be moved from one network to another.
The constraint is rather severe since bridge capacities are not that
high when compared to TVL amounts.

Blockchain funds will calculate portfolio weights and rebalance positions
periodically, - like traditional funds - but with additional decentralization
specific constraints (Donohue \& Yip 2003; Tokat \& Wicas 2007; Calvet
et al., 2009; End-note \ref{enu:Decentralized-finance}). Weight calculations
and rebalancing mechanisms work best with more assets and frequent
rebalancing - both in decentralized and traditional finance. 

Kashyap (2022) describes a range based model tailored to overcome
blockchain nuances such as: gas costs, bridge bottlenecks and higher
volatility and market impact. Each asset will have a range - minimum
and maximum capacity - in terms of the fraction of the TVL allocated
to it. The fundamental idea is to overcome frictions by adjusting
the weight range across the assets. The higher the frictions on an
asset, the wider the range has to be. When there are network wide
constraints - such as the bridge limit - we need to increase the range
on the entire set of assets on the corresponding network. This gives
rise to the aqueduct feature whose capacity for fund flows varies
with constraints.

One option is to have network portfolios - and optimize asset weights
on each network - reducing the need for fund flows across platforms
over a bridge. With sufficient capacity - more assets - on each side
of the bridge and a higher tolerance level - higher range - there
is less need to cross the bridge. If we restrict the funds collected
on one network to that chain alone, we need to ensure that there are
a good number of assets, as part of our portfolio on that network,
to which we can deploy funds. We will have several independent network
portfolios and we have to ensure that no assets get overrepresented
when we aggregate across all the networks. 

This approach is also advisable since at present, “Bridges” built
between various networks are both a “Bottleneck” and an “Achilles
Heel”. Bridges restrict the amount of assets that can flow from one
network to another and they also are vulnerable points for hackers
to target (Li et al., 2020; Lee et al., 2022; Scharfman 2023). Hence
the use of bridges should be cautious initially and depending on asset
flow necessities - plus improvements to the corresponding infrastructure
- we can readjust fund transfer limits. 

Having only network portfolios can mitigate the need for fund flows,
but is unlikely to eliminate it. Another alternative is to have a
global portfolio and from the global weights we can arrive at the
network weights, depending on the restrictions for fund mobility.
For assets that are present on multiple networks, the weights of the
asset on each network have to depend on the proportion of the TVL
on that network in comparison to the overall TVL (End-note \ref{enu:Networks-Rebalancing}).
This methodology involves more complexities - in terms of maintaining
the model range - which we discuss in Section (\ref{sec:Transferring-Assets-Across-Networks-Algorithm}).

\section{\label{sec:Transferring-Assets-Across-Networks-Algorithm}Transferring
Assets Across Networks P,Q}

The algorithm to mitigate bridge limitations is given below, with
clarifications regarding fundamental aspects that motivate various
steps. Based on these factors, we develop detailed formulae that would
help with decision making related to bridge transfers. 

Weight calculations and portfolio rebalancing will be performed periodically
- perhaps at irregular intervals to avoid front running (Bernhardt
\& Taub 2008; Baum et al., 2022; End-note \ref{enu:Front-running,-also}).
When rebalancing happens, new deposits and redemption requests from
investors are actioned. The utilized bridge capacity will depend on
the deployment amount - net of deposits and withdrawals - at that
time in comparison to the total amount already deployed - invested
- on that network. The asset allocation difference between the global
portfolio (across all networks) and the network portfolio will play
a smaller role if each network collects whatever it is able to deploy
within the assets it holds. Such self sufficiency between networks
leads to less transfers. 
\begin{itemize}
\item Consider two platforms $P,Q$ to illustrate our method. For all variables,
the suffixes represent: $P$, $Q$ - network under consideration,
$t$ - current time when observations are made, and $i$ - an asset
counter. The bridge capacity between $P$ and $Q$ is positive: $bridgeCapacity_{PQt}\geq0$.
The capacity from $P$ to $Q$ , $bridgeCapacity_{PQt}$, might be
different from the capacity from $Q$ to $P$, $bridgeCapacity_{QPt}$.
This distinction will be necessary and used accordingly in Step (\ref{enu:Transfer-Amounts-P,Q}). 
\item If we have multiple wallets, $W_{t}$, the $bridgeCapacity_{PQt}$
is the cumulative amount across all the wallets.
\item An asset specific platform indicator, $networkIND_{iPt}=1$, highlights
that asset $i$ is available on network $P$ at time $t$. Otherwise,
$networkIND_{iPt}=0$. 
\end{itemize}
\begin{lyxalgorithm}
When there are bridge constraints, the following algorithm calculates
the bridge transfer amount, the deployment amount across each platform
and the network specific assets weights.
\end{lyxalgorithm}
\begin{enumerate}
\item A separate calculation engine outputs global portfolio weights after
suitable fine tuning. 
\begin{enumerate}
\item $\left(0\leq rminw_{it}\leq ridealw_{it}\leq rmaxw_{it}\right)$ represent
minimum, ideal and maximum raw weights for asset $i$. 
\item Weights are positive $\left(0\leq rminw_{it}\leq ridealw_{it}\leq rmaxw_{it}\right)$
for simplicity, since shorting scenarios are straightforward extensions.
\end{enumerate}
\item Raw weight ranges are extended depending on amounts collected between
rebalance events and total amounts deployed on each network. 
\begin{enumerate}
\item $collectDeployDiff_{PQt+T}$ is an intermediate variable which measures
the difference in amounts collected and amounts to be deployed on
the networks at a future time period, $t+T$, with the information
set given up-to time $t$, $INFO_{t}$. The greater this difference,
the greater the funds that need to move across the bridge. 
\begin{align}
E\left[\left.collectDeployDiff_{PQt+T}\right|INFO_{t}\right] & =\left|\frac{E\left[\left.TBDAmount_{Pt+T}\right|INFO_{t}\right]}{E\left[\left.currentTotalAmount_{Pt+T}\right|INFO_{t}\right]}\right.\nonumber \\
 & -\left.\frac{E\left[\left.TBDAmount_{Qt+T}\right|INFO_{t}\right]}{E\left[\left.currentTotalAmount_{Qt+T}\right|INFO_{t}\right]}\right|\label{eq:Collect-Deploy-Diff}
\end{align}
\item $TBDAmount_{Pt}$ and $TBDAmount_{Qt}$ are net new amounts to be
deployed - which are funds accumulated since the last rebalancing
event; $currentTotalAmount_{Pt}$ and $currentTotalAmount_{Qt}$ are
notional amounts invested across all existing assets on networks $P$
and $Q$. $TBDAmount_{Pt,Qt,}$ can be positive or negative depending
on whether we have a net deposit or withdrawal scenario. Since shorting
is not allowed, $currentTotalAmount_{Pt,Qt}\geq0$ . The following
condition - total withdrawal has to be less than total current investment
- will be satisfied,
\begin{align}
\left(-1\right)*\left\{ E\left[\left.TBDAmount_{Pt+T}\right|INFO_{t}\right]+E\left[\left.TBDAmount_{Qt+T}\right|INFO_{t}\right]\right\} \nonumber \\
\leq E\left[\left.currentTotalAmount_{Pt+T}\right|INFO_{t}\right]+E\left[\left.currentTotalAmount_{Qt+T}\right|INFO_{t}\right]\label{eq:Net-Deployment-Less-Investment}
\end{align}
\item The current notional amounts invested across all assets on a network,
$currentTotalAmount_{t}$, is the sum of the amount in each asset
$i$, $currentAmount_{it}$, which is quantity $q_{it}$ of the asset
times its latest price $p_{it}$.
\item \label{enu:Transfer-Need}$\left[\frac{E\left[\left.TBDAmount_{Pt+T}\right|INFO_{t}\right]}{E\left[\left.currentTotalAmount_{Pt+T}\right|INFO_{t}\right]}\right]$
is the ratio of amount to be deployed by amount invested on a network.
Higher positive values indicate greater need for fund outflows and
vice versa.
\item $E\left[\left.\cdots\right|INFO_{t}\right]$ denotes the expectation
operator based on the information set given up-to time $t$, $INFO_{t}$.
Historical average values can serve as proxies for future expected
values. For example, the below historical average is an approximation
for $E\left[\left.TBDAmount_{Pt+T}\right|INFO_{t}\right]$. 
\begin{align}
E\left[\left.TBDAmount_{Pt+T}\right|INFO_{t}\right] & \approx\underset{n\rightarrow\infty}{\lim}\left[\frac{1}{n}\sum_{i=1}^{n}TBDAmount_{P\left\{ t-T_{1}+\left(i-1\right)\left(\Delta t\right)\right\} ,\left\{ t-T_{1}+i\left(\Delta t\right)\right\} }\right]\\
\Delta t & =\frac{T_{1}}{n}
\end{align}
$T_{1}$ is a sufficiently long time period. Hence $t-T_{1}$ is a
historical time period going back from $t$ towards the time since
the system has been operating. $\Delta t\approx T$ is the duration
for which we are making the forecast. $TBDAmount_{P\left\{ t_{1}\right\} ,\left\{ t_{2}\right\} }$
is the change in the variable from $t_{1}$ to $t_{2}$. Care needs
to be taken to exclude - or handle accordingly - the drastic changes
in the variables when rebalancing happens. Clearly numerous alternate
formulations - including different weights for different intervals
in the historical average and so on - are possible. 
\item Another alternative is to forecast the component variables - fund
quantity, asset quantities, fund price, asset prices, cash - separately
and aggregate them to get the amounts currently invested and to be
deployed. These variables take only positive values - if redemption
related variables are handled properly - and can be forecast using
Geometric Brownian Motions (GBMs: End-note \ref{enu:Numerous-alternate-Formulations}). 
\item We can also consider actual observations since the last rebalancing
event, $t-T_{2}$, to the current time, $t$, right now when we need
to perform a rebalancing event. These actual values can be used to
arrive at the bridge transfer amounts. In such a case we do not need
to forecast any variables but instead we will be using actual changes
observed during the time period since the last rebalancing event.
Notice that forecasts can be helpful since the predicted variables
can help to plan future rebalancing events by indicating whether there
will be a need to perform a rebalancing event sooner than anticipated.
\end{enumerate}
\item The percentage to extend the weights, $bridgeStretch_{PQt+T}$, based
on the bridge constraints is calculated in the below formula:
\begin{align}
E\left[\left.bridgeStretch_{PQt+T}\right|INFO_{t}\right] & =E\left[\left.collectDeployDiff_{PQt+T}\right|INFO_{t}\right]\nonumber \\
 & \left(1+\left\{ \frac{NUMER.EXPR}{DENOM.EXPR}\right\} \right)\label{eq:Bridge-Stretch}
\end{align}
\begin{align}
NUMER.EXPR & =E\left[\left.TBDAmount_{Pt+T}\right|INFO_{t}\right]+E\left[\left.TBDAmount_{Qt+T}\right|INFO_{t}\right]\\
DENOM.EXPR & =E\left[\left.currentTotalAmount_{Pt+T}\right|INFO_{t}\right]\nonumber \\
 & +E\left[\left.currentTotalAmount_{Qt+T}\right|INFO_{t}\right]+E\left[\left.bridgeCapacity_{PQt+T}\right|INFO_{t}\right]
\end{align}

\begin{enumerate}
\item $E\left[\left.bridgeStretch_{PQt+T}\right|INFO_{t}\right]$ will be
a positive number - using (Eq: \ref{eq:Net-Deployment-Less-Investment})
- since, 
\begin{equation}
-1<\frac{E\left[\left.TBDAmount_{Pt+T}\right|INFO_{t}\right]+E\left[\left.TBDAmount_{Qt+T}\right|INFO_{t}\right]}{E\left[\left.currentTotalAmount_{Pt+T}+currentTotalAmount_{Qt+T}+bridgeCapacity_{PQt+T}\right|INFO_{t}\right]}\leq\infty
\end{equation}
If $E\left[\left.TBDAmount_{Pt+T}\right|INFO_{t}\right]<0$ or $E\left[\left.TBDAmount_{Qt+T}\right|INFO_{t}\right]<0$
then,
\begin{align}
\left|E\left[\left.TBDAmount_{Pt+T}\right|INFO_{t}\right]\right| & \leq E\left[\left.currentTotalAmount_{Pt+T}\right|INFO_{t}\right]\nonumber \\
 & +E\left[\left.currentTotalAmount_{Qt+T}\right|INFO_{t}\right]\label{eq:TBD-Condition-P}\\
OR\nonumber \\
\left|E\left[\left.TBDAmount_{Qt+T}\right|INFO_{t}\right]\right| & \leq E\left[\left.currentTotalAmount_{Pt+T}\right|INFO_{t}\right]\nonumber \\
 & +E\left[\left.currentTotalAmount_{Qt+T}\right|INFO_{t}\right]\label{eq:TBD-Condition-Q}
\end{align}
If $E\left[\left.TBDAmount_{Pt+T}\right|INFO_{t}\right]<0$ and $E\left[\left.TBDAmount_{Qt+T}\right|INFO_{t}\right]<0$
then,
\begin{align}
\left|E\left[\left.TBDAmount_{Pt+T}\right|INFO_{t}\right]+E\left[\left.TBDAmount_{Qt+T}\right|INFO_{t}\right]\right| & \leq\left|E\left[\left.currentTotalAmount_{Pt+T}\right|INFO_{t}\right]\right.\nonumber \\
 & +E\left[\left.currentTotalAmount_{Qt+T}\right|INFO_{t}\right]\nonumber \\
 & +\left.E\left[\left.bridgeCapacity_{PQt+T}\right|INFO_{t}\right]\right|
\end{align}
\item It is sensible to restrict the maximum bridge stretch value used -
to be practical when huge amounts are being deposited or withdrawn
- to extend the weight range, $bridSTRCH_{t}$, as follows,
\begin{equation}
bridSTRCH_{t}=\max\left(\left|E\left[\left.bridgeStretch_{PQt+T}\right|INFO_{t}\right]\right|,MAXBRIDGESTRETCH_{t}\right)
\end{equation}
A recommended value for $MAXBRIDGESTRETCH_{t}=0.2$, stretching bridge
usage by a maximum of 20\%. Clearly, alternative values have to be
figured out based on the specifics of the usage scenarios.
\end{enumerate}
\item The raw asset weights will be extended using $bridSTRCH_{t}$ as follows,
\begin{align}
minw_{it} & =\left(rminw_{it}\right)\left(1-bridSTRCH_{t}\right)\\
maxw_{it} & =\left(rmaxw_{it}\right)\left(1+bridSTRCH_{t}\right)
\end{align}
$\left(minw_{it},idealw_{it},maxw_{it}\right)$ are minimum, ideal
and maximum weights for asset $i$ after incorporating bridge constraints. 
\begin{enumerate}
\item The ideal weight - which is unaltered - is a good risk management
reference point.
\end{enumerate}
\item \label{enu:Asset-Network-Weights}The stretched asset weights will
be adjusted to network specific weights, $\left(minw_{iPt},idealw_{iPt},maxw_{iPt}\right)$,
for assets that are available on both networks. This modification
is based on the proportion of TVL that each network will hold after
including the net new amount to be deployed or withdrawn. 
\begin{align}
minw_{iPt} & =\left(minw_{it}\right)\left(networkWeight_{iPt}\right)\\
idealw_{iPt} & =\left(idealw_{it}\right)\left(networkWeight_{iPt}\right)\\
maxw_{iPt} & =\left(rmaxw_{it}\right)\left(networkWeight_{iPt}\right)
\end{align}
\begin{align}
networkWeight_{iPt} & =\left(\frac{NUMER.EXPRTWO}{DENOM.EXPRTWO}\right)\\
NUMER.EXPRTWO & =\left(networkIND_{iPt}\right)\left(TBDAmount_{Pt}+currentTotalAmount_{Pt}\right)\\
DENOM.EXPRTWO & =\left(networkIND_{iPt}\right)\left(TBDAmount_{Pt}+currentTotalAmount_{Pt}\right)\nonumber \\
 & +\left(networkIND_{iQt}\right)\left(TBDAmount_{Qt}+currentTotalAmount_{Qt}\right)
\end{align}

\begin{enumerate}
\item For assets available on only one network the stretched weights are
unaltered. The network indicator ensures that assets not present on
a network have weight zero.
\item \label{enu:The-stretched-weights-restricted}The stretched weights
can be restricted to be within certain bounds so that we stay aligned
with portfolio risk and return preferences. The below formulation
ensures that the weights satisfy no shorting criteria when $MINWEIGHT_{iPt}=0$
and $MAXWEIGHT_{iPt}=1$.
\begin{align}
networkTrimWeight_{iPt} & =\min\left[\max\left(networkWeight_{iPt},MINWEIGHT_{iPt}\right),MAXWEIGHT_{iPt}\right]\nonumber \\
 & +\max\left[\min\left(networkWeight_{iPt},MAXWEIGHT_{iPt}\right),MINWEIGHT_{iPt}\right]\label{eq:Network-Weight-P}
\end{align}
\begin{align}
networkTrimWeight_{iQt} & =\left(1-networkTrimWeight_{iPt}\right)\label{eq:Network-Weight-Q}
\end{align}
\end{enumerate}
\item \label{enu:min-max-capacity}Calculate the minimum and maximum total
capacity, $minNetworkCapacity_{Pt}$, $minNetworkCapacity_{Qt}$,
$maxNetworkCapacity_{Pt}$, $maxNetworkCapacity_{Qt}$ on networks
$P,Q$ respectively. These represent the minimum or maximum band for
the total investment on that network.
\begin{equation}
minNetworkCapacity_{Pt}=\left(currTotalPlusTBDAmount_{PQt}\right)\left(\sum_{i=1}^{k_{Pt}}minw_{iPt}\right)\label{eq:minNetworkCapacity_P}
\end{equation}
\begin{equation}
minNetworkCapacity_{Qt}=\left(currTotalPlusTBDAmount_{PQt}\right)\left(\sum_{i=1}^{k_{Qt}}minw_{iQt}\right)\label{eq:minNetworkCapacity_Q}
\end{equation}
\begin{equation}
maxNetworkCapacity_{Pt}=\left(currTotalPlusTBDAmount_{PQt}\right)\left(\sum_{i=1}^{k_{Pt}}maxw_{iPt}\right)\label{eq:maxNetworkCapacity_P}
\end{equation}
\begin{equation}
maxNetworkCapacity_{Qt}=\left(currTotalPlusTBDAmount_{PQt}\right)\left(\sum_{i=1}^{k_{Qt}}maxw_{iQt}\right)\label{eq:maxNetworkCapacity_Q}
\end{equation}
\begin{align}
currTotalPlusTBDAmount_{PQt} & =currentTotalAmount_{Pt}+currentTotalAmount_{Qt}\nonumber \\
 & +TBDAmount_{Pt}+TBDAmount_{Qt}
\end{align}
\begin{align}
currTotalPlusTBDAmount_{Pt} & =currentTotalAmount_{Pt}+TBDAmount_{Pt}\label{eq:Current-Plus-TBD-P}
\end{align}
\begin{align}
currTotalPlusTBDAmount_{Qt} & =currentTotalAmount_{Qt}+TBDAmount_{Qt}\label{eq:Current-Plus-TBD-Q}
\end{align}
$currTotalPlusTBDAmount_{PQt}$ represents the total amount that will
be held across both networks; $currTotalPlusTBDAmount_{Pt}$, $currTotalPlusTBDAmount_{Qt}$
represent the amounts that will be held across networks, $P,Q$;$k_{Pt}$,
$k_{Qt}$ are the total number of assets on platforms $P,Q$. The
total amount held is the sum of the amounts already invested plus
net deposits or withdrawals after the latest rebalancing.
\begin{enumerate}
\item Due to altering of stretched weights in Step (\ref{enu:The-stretched-weights-restricted}),
we aggregate allocated amounts based on trimmed asset weights - rather
than aggregating stretched weights and multiplying the total amount. 
\end{enumerate}
\item Calculate the difference between amounts that will be deployed across
each network - current plus net new funds: $currTotalPlusTBDAmount_{Pt}$,
$currTotalPlusTBDAmount_{Qt}$ - with the minimum and maximum capacity
on that network. If the total amount is below the minimum or above
the maximum capacity then funds will need to be received from or moved
to the other network respectively. 
\begin{align}
amountOutsideBand_{Pt} & =\min\left[\left(currTotalPlusTBDAmount_{Pt}-minNetworkCapacity_{Pt}\right),0\right]\nonumber \\
 & +\max\left[\left(currTotalPlusTBDAmount_{Pt}-maxNetworkCapacity_{Pt}\right),0\right]\label{eq:Outside-Band-P}
\end{align}
\begin{align}
amountOutsideBand_{Qt} & =\min\left[\left(currTotalPlusTBDAmount_{Qt}-minNetworkCapacity_{Qt}\right),0\right]\nonumber \\
 & +\max\left[\left(currTotalPlusTBDAmount_{Qt}-maxNetworkCapacity_{Qt}\right),0\right]\label{eq:Outside-Band-Q}
\end{align}
\begin{align}
maxSend_{Pt} & =\max\left[\left(currTotalPlusTBDAmount_{Pt}-minNetworkCapacity_{Pt}\right),0\right]\label{eq:max-send-P}
\end{align}
\begin{align}
maxSend_{Qt} & =\max\left[\left(currTotalPlusTBDAmount_{Qt}-minNetworkCapacity_{Qt}\right),0\right]\label{eq:max-send-Q}
\end{align}
\begin{align}
maxRecieve_{Pt} & =\max\left[\left(maxNetworkCapacity_{Pt}-currTotalPlusTBDAmount_{Pt}\right),0\right]\label{eq:max-receive-P}
\end{align}
\begin{align}
maxRecieve_{Qt} & =\max\left[\left(maxNetworkCapacity_{Qt}-currTotalPlusTBDAmount_{Qt}\right),0\right]\label{eq:max-receive-Q}
\end{align}

\begin{enumerate}
\item If $amountOutsideBand_{Pt}$, $amountOutsideBand_{Qt}$ - amounts
outside the minimum and maximum bands on networks, $P,Q$ - are negative
then the corresponding network has to receive funds and vice versa
for positive values.
\item $maxSend_{Pt},maxRecieve_{Pt},maxSend_{Qt},maxRecieve_{Qt}$are maximum
amounts that can be sent to or received from $P,Q$ respectively.
\end{enumerate}
\item \label{enu:Transfer-Amounts-P,Q}Calculate the amount to be transferred
from $P$ to $Q$, $transferAmount_{PQt}$ or from $Q$ to $P$, $transferAmount_{QPt}$.
This is the amount outside the maximum or minimum bands compared with
what the capacity to receive or send on the other network.
\begin{align}
transferAmount_{PQt} & =\min\left[\max\left\{ \min\left(amountOutsideBand_{Pt},maxRecieve_{Qt}\right),0\right\} ,bridgeCapacity_{PQt}\right]\nonumber \\
+ & \max\left[\min\left\{ \max\left(amountOutsideBand_{Pt},\left(-1\right)*maxSend_{Qt}\right),0\right\} ,\left(-1\right)*bridgeCapacity_{QPt}\right]\label{eq:Transfer-Simple-PQ}
\end{align}
\begin{align}
transferAmount_{QPt} & =\min\left[\max\left\{ \min\left(amountOutsideBand_{Qt},maxRecieve_{Pt}\right),0\right\} ,bridgeCapacity_{QPt}\right]\nonumber \\
+ & \max\left[\min\left\{ \max\left(amountOutsideBand_{Qt},\left(-1\right)*maxSend_{Pt}\right),0\right\} ,\left(-1\right)*bridgeCapacity_{PQt}\right]\label{eq:Transfer-Simple-QP}
\end{align}

\begin{enumerate}
\item $transferAmount_{PQt}$ and $transferAmount_{QPt}$ will both be zero
or one will be positive and the other negative of equal magnitude
under most circumstances. 
\item When the amount above the maximum band in one network is less than
the amount needed on the other network - to meet withdrawal requests
- then one of the transfer amounts will be negative and larger than
the other. In this case the full redemption request cannot be met
and multiple transfers are needed to fulfill the entire withdraw amount. 
\item The below alternate formulation can be used so that we can meet redemption
requests as best as possible without waiting for later rounds,
\begin{align}
transferAmount_{PQt} & =\min\left[\max\left\{ \min\left(\max\left[\vphantom{\left[\frac{\left(amountOutsideBand_{Pt}+\triangle\right)}{\left|amountOutsideBand_{Pt}\right|+\triangle}\right]}amountOutsideBand_{Pt},\left(-1\right)*amountOutsideBand_{Qt}\right]\right.\right.\right.\nonumber \\
 & *\left[\frac{\max\left(amountOutsideBand_{Pt}+\triangle,0\right)}{\left|amountOutsideBand_{Pt}\right|+\triangle}\right]\nonumber \\
 & \left.\left.\left.\vphantom{\left[\frac{\left(amountOutsideBand_{Pt}+\triangle\right)}{\left|amountOutsideBand_{Pt}\right|+\triangle}\right]},maxRecieve_{Qt},maxSend_{Pt}\right),0\right\} ,bridgeCapacity_{PQt}\right]\\
+ & \max\left[\min\left\{ \max\left(\vphantom{\left[\frac{\left(amountOutsideBand_{Pt}+\triangle\right)}{\left|amountOutsideBand_{Pt}\right|+\triangle}\right]}amountOutsideBand_{Pt},\right.\right.\right.\nonumber \\
 & \left.\left.\left.\vphantom{\left[\frac{\left(amountOutsideBand_{Pt}+\triangle\right)}{\left|amountOutsideBand_{Pt}\right|+\triangle}\right]}\left(-1\right)*maxSend_{Qt}\right),0\right\} ,\left(-1\right)*bridgeCapacity_{QPt}\right]\label{eq:Alternate-PQ}
\end{align}
\begin{align}
transferAmount_{QPt} & =\min\left[\max\left\{ \min\left(\max\left[\vphantom{\left[\frac{\left(amountOutsideBand_{Pt}+\triangle\right)}{\left|amountOutsideBand_{Pt}\right|+\triangle}\right]}amountOutsideBand_{Qt},\left(-1\right)*amountOutsideBand_{Pt}\right]\right.\right.\right.\nonumber \\
 & *\left[\frac{\max\left(amountOutsideBand_{Qt}+\triangle,0\right)}{\left|amountOutsideBand_{Qt}\right|+\triangle}\right]\nonumber \\
 & \left.\left.\left.\vphantom{\left[\frac{\left(amountOutsideBand_{Pt}+\triangle\right)}{\left|amountOutsideBand_{Pt}\right|+\triangle}\right]},maxRecieve_{Pt},maxSend_{Qt}\right),0\right\} ,bridgeCapacity_{QPt}\right]\\
+ & \max\left[\min\left\{ \max\left(\vphantom{\left[\frac{\left(amountOutsideBand_{Pt}+\triangle\right)}{\left|amountOutsideBand_{Pt}\right|+\triangle}\right]}amountOutsideBand_{Qt},\right.\right.\right.\nonumber \\
 & \left.\left.\left.\vphantom{\left[\frac{\left(amountOutsideBand_{Pt}+\triangle\right)}{\left|amountOutsideBand_{Pt}\right|+\triangle}\right]}\left(-1\right)*maxSend_{Pt}\right),0\right\} ,\left(-1\right)*bridgeCapacity_{PQt}\right]\label{eq:Alternate-QP}
\end{align}
$\triangle$ is a small positive value lesser than the smallest allowed
withdrawal request amount (End-note \ref{enu:Delta-value-of}).
\item Transfers in only one direction are needed, if the implementation
is precise and weights are proper - otherwise netting the amounts
might be necessary. Careful execution is needed for scenarios such
as: capacity range is not sufficiently wide; minimum weights on one
network are above maximum weights of the other; discrepancies between
relative levels of deployed amounts and new deposits or withdrawals;
other related reasons (Section \ref{sec:Conclusion}).
\item Figures (\ref{fig:Simulation-Parameters}; \ref{fig:External-Inputs-Systems};
\ref{fig:Transfer-Amounts-Primary}; \ref{fig:Intermediate-State-Variables})
in Section (\ref{sec:Numerical-Illustrations}) provide numerical
examples and illustrate cases with different weights, current investment
amounts, deposits, withdrawals and bridge capacities.
\end{enumerate}
\end{enumerate}

\section{\label{sec:Numerical-Illustrations}Numerical Illustrations}

Each of the tables in this section is relevant to the algorithm described
in Section (\ref{sec:Transferring-Assets-Across-Networks-Algorithm})
and are referenced in the steps of the algorithm. The tables provide
many scenarios related to the algorithm indicating the amounts that
need to be transferred across the networks. 

The inputs and outputs to the system are given and clearly explained.
The numerical scenarios we have provided use simulated data. The parameters
used for the simulations are given separately as are several intermediate
variables that are necessary to arrive at the outputs. The outputs
and the intermediate variables can be helpful to understand how the
system is functioning. More detailed data from the simulations can
be provided to the readers upon request.

\begin{figure}[H]
\includegraphics[width=9cm]{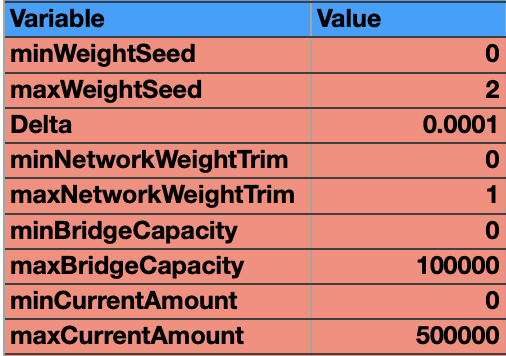}

\caption{\label{fig:Simulation-Parameters}Simulation Parameters}
\end{figure}

Below, we provide supplementary descriptions for each table. These
descriptions help the readers to better understand the elaborate explanations
we have given in the main text for the mathematical equations and
conditions for each step of the blockchain bridge algorithm. 
\begin{itemize}
\item The Table in Figure (\ref{fig:Simulation-Parameters}) shows the parameters
we have used to simulate global weights, bridge capacities, current
invested amounts, deposit and withdrawals from uniform random distributions
(End-note \ref{enu:Uniform-Distribution}). We have used uniform distributions
for simplicity since the main goal of these numerical results are
to check whether the bridge algorithm is working effectively under
several different scenarios. We need to seed uniform distributions
with a minimum and maximum value which are given below for the different
variables.
\item The rows in Figure (\ref{fig:Simulation-Parameters}) represent the
following information respectively: 
\begin{enumerate}
\item \textbf{minWeightSeed }is the minimum value of the random variable
for the global weights. The global weights are randomly chosen between
the \textbf{minWeightSeed }and\textbf{ maxWeightSeed. }
\item \textbf{maxWeightSeed }is the maximum value of the random variable
for the global weights. 
\begin{enumerate}
\item For the minimum global weights we randomly sample a value from a uniform
distribution with lower and upper bounds given by \textbf{minWeightSeed
}and\textbf{ maxWeightSeed. 
\begin{equation}
minWeight_{it}\sim U\left[minWeightSeed,maxWeightSeed\right]
\end{equation}
}Here, $U$ stands for uniform distribution.
\item For the maximum global weights we sample a value between the minimum
global weight for that scenario and the upper bound \textbf{maxWeightSeed.
\begin{equation}
maxWeight_{it}\sim U\left[minWeight_{it}maxWeightSeed\right]
\end{equation}
}
\end{enumerate}
\item \textbf{Delta represents $\triangle$ }in Equations (\ref{eq:Alternate-PQ};
\ref{eq:Alternate-QP}), which is a small positive value lesser than
the smallest allowed withdrawal request amount.
\item \textbf{minNetworkWeightTrim }represents $MINWEIGHT_{iPt}$, which
is the minimum value of the network weights in Equation (\ref{eq:Network-Weight-P};
\ref{eq:Network-Weight-Q}). Values below \textbf{minNetworkWeightTrim}
are set to this value. 
\item \textbf{maxNetworkWeightTrim }represents $MAXWEIGHT_{iPt}$, which
is the maximum value of the network weights in Equation (\ref{eq:Network-Weight-P};
\ref{eq:Network-Weight-Q}). Values above \textbf{maxNetworkWeightTrim}
are set to this value. 
\item \textbf{minBridgeCapacity }is the minimum value of the random variable
for the bridge capacity, $bridgeCapacity_{PQt,QPt}$, in United States
Dollars (USD). 
\item \textbf{maxBridgeCapacity} is the maximum value of the random variable
for the bridge capacity in United States Dollars (USD). 
\begin{enumerate}
\item The bridge capacity is randomly chosen from a uniform distribution
with lower and upper bounds given by \textbf{minBridgeCapacity }and\textbf{
maxBridgeCapacity.}
\end{enumerate}
\item \textbf{minCurrentAmount }is the minimum value of the random variable
current invested amount, $currentTotalAmount_{Pt,Qt}$, in United
States Dollars (USD). 
\item \textbf{maxCurrentAmount} is the maximum value of the random variable
current invested amount in United States Dollars (USD). 
\begin{enumerate}
\item The current invested amount is randomly chosen from a uniform distribution
with lower and upper bounds given by \textbf{minCurrentAmount }and\textbf{
maxCurrentAmount.} 
\item The amounts to be deployed, $TBDAmount_{Pt,Qt}$, are chosen such
that Equations (\ref{eq:Net-Deployment-Less-Investment}; \ref{eq:TBD-Condition-P};
\ref{eq:TBD-Condition-Q}) are satisfied. 
\item The amount to be deployed on one network is selected from a uniform
distribution with lower bound that is the sum of the current invested
amounts on both networks with a negative sign. This indicates that
the largest amount that can be withdrawn is the sum of the total invested
on both networks. 
\begin{align}
TBDAmount_{Pt} & \sim U\left[\left(-1\right)*\left(currentTotalAmount_{Pt}+currentTotalAmount_{Qt}\right)\right.\\
 & \left.,maxCurrentAmount\right]
\end{align}
\item The amount to be deployed on the second network is selected from a
uniform distribution with lower bound that depends on whether the
first network has a net withdrawal case or not. If the first network
has a net withdrawal scenario then the lower bound for the second
network will be the sum of the current invested amounts on both networks
with a negative sign less the withdrawal amount on the first network.
If the first network has a net deposit scenario then the lower bound
for the second network will be the sum of the current invested amounts
on both networks with a negative sign.
\begin{align}
TBDAmount_{Qt} & \sim U\left[\left(-1\right)*\left(currentTotalAmount_{Pt}+currentTotalAmount_{Qt}\right)\right.\\
 & +\max\left\{ \left(-1\right)*TBDAmount_{Pt},0\right\} \left.,maxCurrentAmount\right]
\end{align}
\item The upper bound for the uniform distribution for both networks is
the \textbf{maxCurrentAmount.}
\end{enumerate}
\end{enumerate}
\end{itemize}

\begin{figure}[H]
\includegraphics[width=18cm]{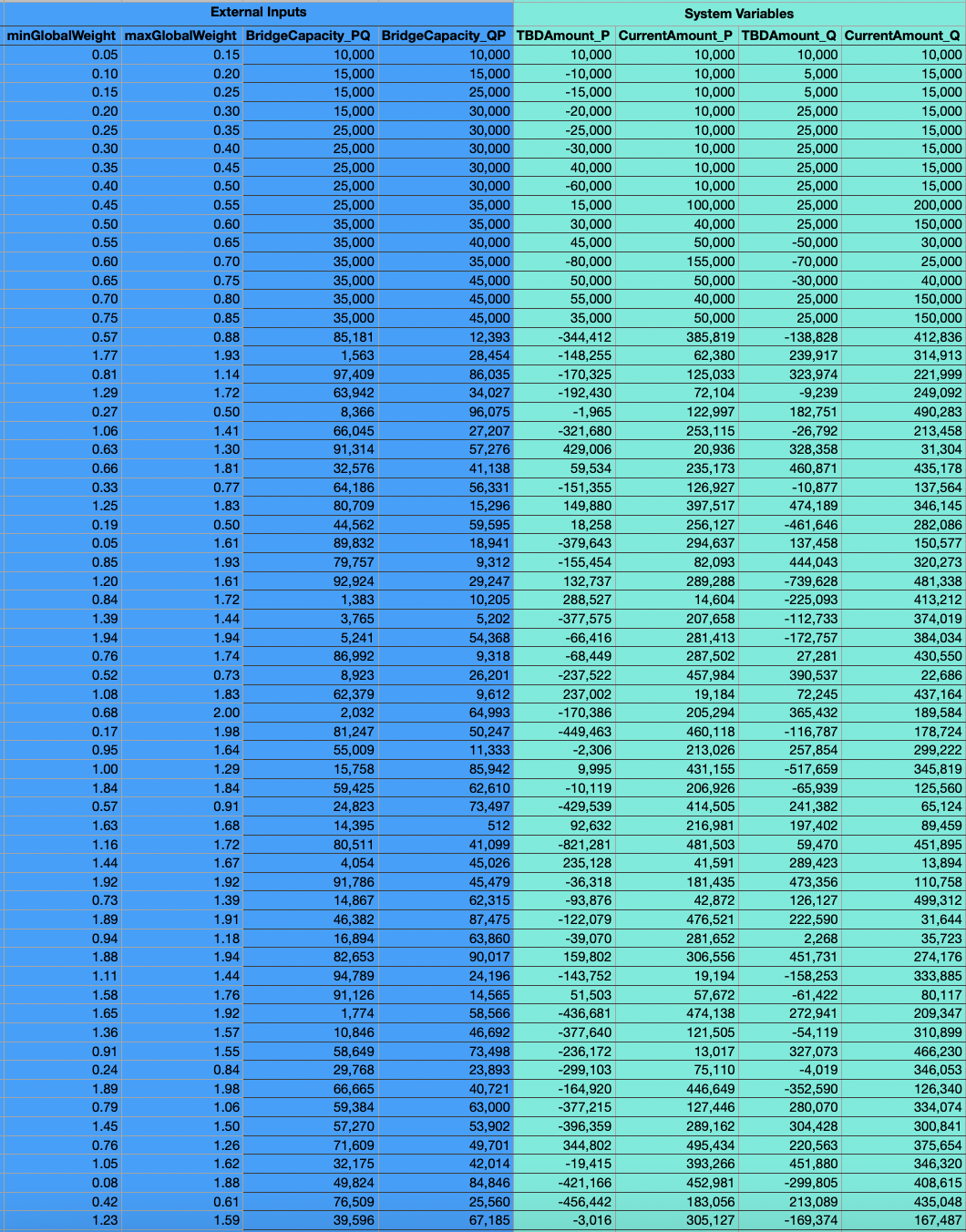}

\caption{\label{fig:External-Inputs-Systems}External Inputs and System Variables}
\end{figure}

\begin{itemize}
\item The Table in Figure (\ref{fig:External-Inputs-Systems}) shows the
external inputs and the system variables which are simulated from
uniform distributions with certain conditions being satisfied as mentioned
in the discussion for Figure (\ref{fig:Simulation-Parameters}). The
first 15 rows represent scenarios based on manually selected values
chosen to illustrate certain special cases. The rest of the rows are
based on values being sampled randomly.
\item The columns in Figure (\ref{fig:External-Inputs-Systems}) - corresponding
to the scenarios in each row - represent the following information
respectively: 
\begin{enumerate}
\item \textbf{minGlobalWeight }is the minimum global weight for that scenario. 
\item \textbf{maxGlobalWeight }is the maximum global weight for that scenario. 
\item \textbf{BridgeCapacity\_PQ }represents bridge capacity from $P$ to
$Q$, $bridgeCapacity_{PQt}$.
\item \textbf{BridgeCapacity\_QP }represents bridge capacity from $Q$ to
$P$, $bridgeCapacity_{QPt}$. 
\item \textbf{TBDAmount\_P }is the amount to be deployed on network $P$,
$TBDAmount_{Pt}$. This is the net of deposits and withdrawals on
network $P$.
\item \textbf{CurrentAmount\_P} is the amount currently invested on network
$P$, $currentTotalAmount_{Pt}.$
\item \textbf{TBDAmount\_Q }is the amount to be deployed on network $Q$,
$TBDAmount_{Qt}$. This is the net of deposits and withdrawals on
network $Q$. 
\item \textbf{CurrentAmount\_Q} is the amount currently invested on network
$Q$, $currentTotalAmount_{Qt}.$
\end{enumerate}
\end{itemize}

\begin{figure}[H]
\includegraphics[width=18cm]{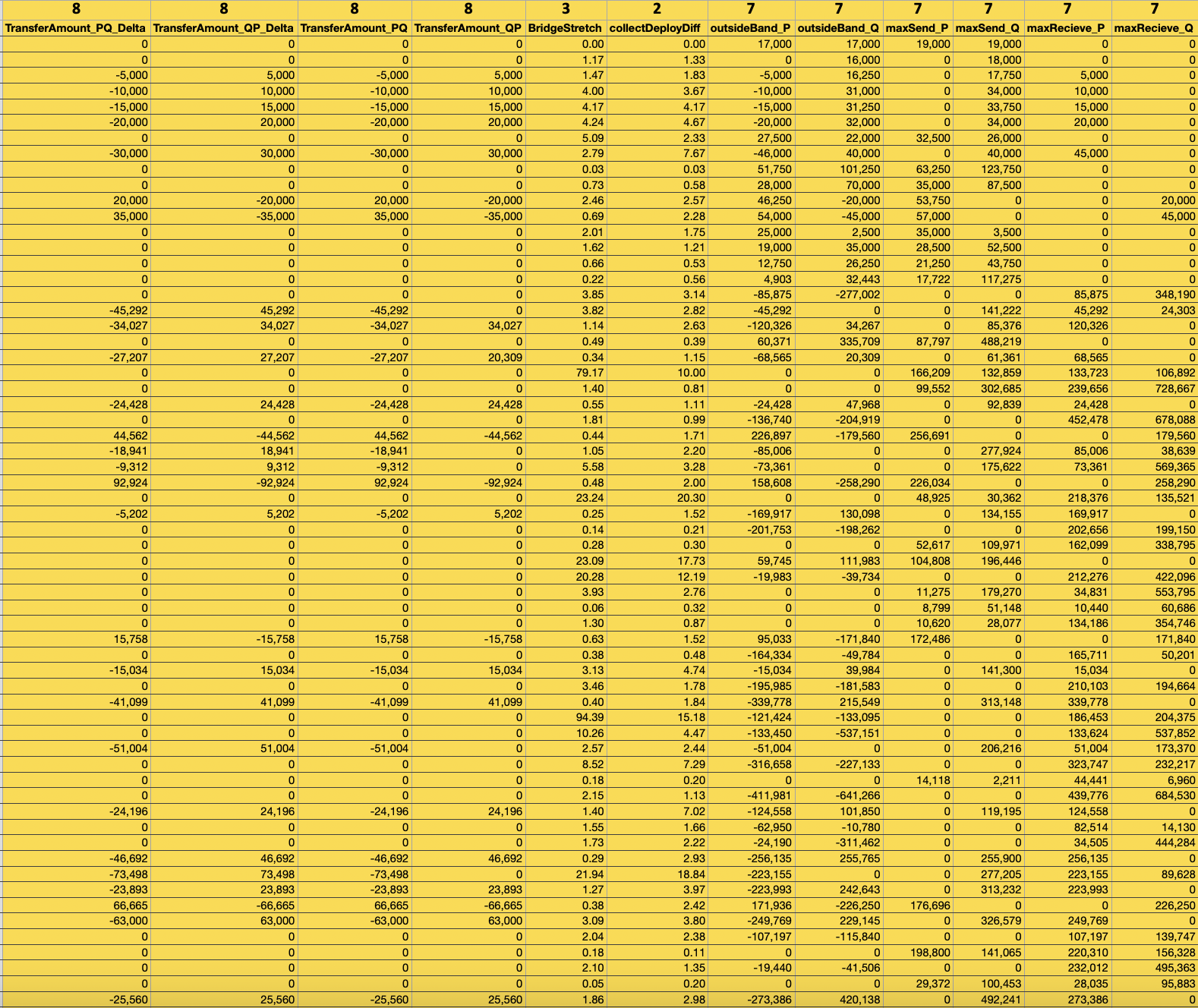}

\caption{\label{fig:Transfer-Amounts-Primary}Transfer Amounts and Primary
Outputs}
\end{figure}

\begin{figure}[H]
\includegraphics[width=18cm]{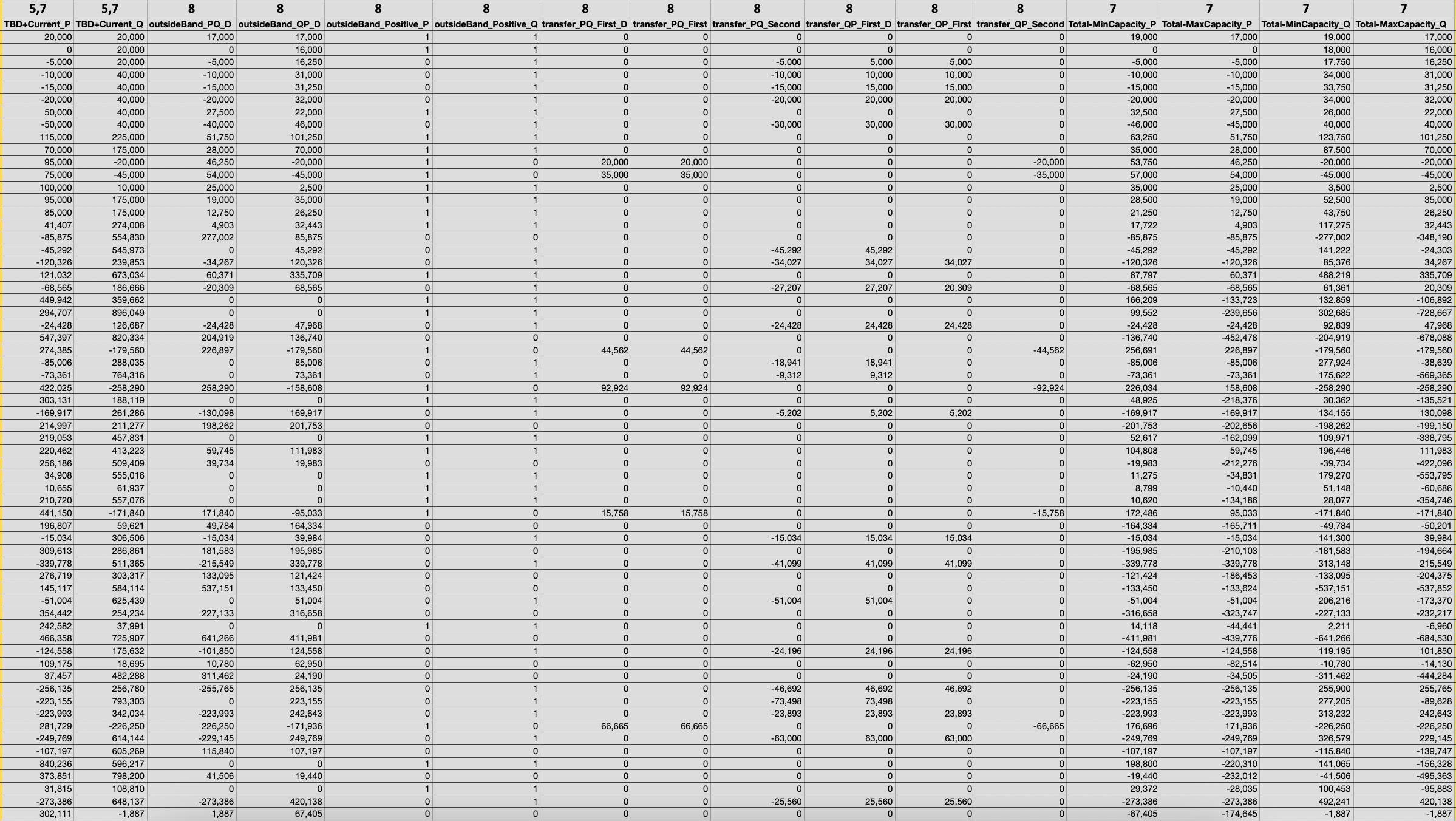}

\caption{\label{fig:Intermediate-State-Variables}Intermediate State Variables}
\end{figure}

\begin{itemize}
\item The Table in Figure (\ref{fig:Transfer-Amounts-Primary}) shows the
amounts to be transferred across the networks and other key outputs
which can help to understand how the system is functioning. The first
15 rows represent scenarios based on manually selected values chosen
to illustrate certain special cases. The rest of the rows are based
on values selected randomly. The scenarios in each row are the same
as those corresponding to the rows given in Figures (\ref{fig:External-Inputs-Systems};
\ref{fig:Intermediate-State-Variables}). The first row indicates
the steps in the algorithm to which each column corresponds to.
\item The columns in Figure (\ref{fig:Transfer-Amounts-Primary}) - corresponding
to the scenarios in each row - represent the following information
respectively: 
\begin{enumerate}
\item \textbf{TransferAmount\_PQ\_Delta }is the amount to be transferred
from $P$ to $Q$ corresponding to the formulation in Equation (\ref{eq:Alternate-PQ}).
\item \textbf{TransferAmount\_QP\_Delta }is the amount to be transferred
from $Q$ to $P$ corresponding to the formulation in Equation (\ref{eq:Alternate-QP}).
\item \textbf{TransferAmount\_PQ }is the amount to be transferred from $P$
to $Q$ corresponding to the formulation in Equation (\ref{eq:Transfer-Simple-PQ}).
\item \textbf{TransferAmount\_QP }is the amount to be transferred from $Q$
to $P$ corresponding to the formulation in Equation (\ref{eq:Transfer-Simple-QP}).
\begin{enumerate}
\item Notice that the \textbf{TransferAmount\_PQ\_Delta }and\textbf{ TransferAmount\_QP\_Delta
}are equal and opposite in magnitude. But\textbf{ TransferAmount\_PQ
}and\textbf{ TransferAmount\_QP }can be different as seen in the scenarios
corresponding to rows 18 and 21. In row 18 and 21, \textbf{TransferAmount\_PQ}
= (-45,292 ; -27,207) and \textbf{TransferAmount\_QP }= (0 ; 20309).
\end{enumerate}
\item \textbf{BridgeStretch }represents $bridgeStretch_{PQt}$ given in
Equation (\ref{eq:Bridge-Stretch}), which is the percentage to extend
the weights based on the bridge constraints. 
\item \textbf{collectDeployDiff} represents $collectDeployDiff_{PQt}$ given
in Equation (\ref{eq:Collect-Deploy-Diff}), which measures the difference
in amounts collected and amounts to be deployed on the networks.
\item \textbf{outsideBand\_P }represents $amountOutsideBand_{Pt}$ given
in Equation (\ref{eq:Outside-Band-P}), which measures amount outside
the minimum and maximum bands - or the amount above or below the minimum
and maximum total capacity - on network $P$.
\item \textbf{outsideBand\_Q} represents $amountOutsideBand_{Qt}$ given
in Equation (\ref{eq:Outside-Band-Q}), which measures amount outside
the minimum and maximum bands - or the amount above or below the minimum
and maximum total capacity - on network $Q$.
\item \textbf{maxSend\_P }represents $maxSend_{Pt}$ given in Equation (\ref{eq:max-send-P}),
which measures the maximum amount that can be sent from network $P$.
\item \textbf{maxSend\_Q }represents $maxSend_{Qt}$ given in Equation (\ref{eq:max-send-Q}),
which measures the maximum amount that can be sent from network $Q$.
\item \textbf{maxRecieve\_P }represents $maxRecieve_{Pt}$ given in Equation
(\ref{eq:max-receive-P}), which measures the maximum amount that
can be received by network $P$.
\item \textbf{maxRecieve\_Q }represents $maxRecieve_{Qt}$ given in Equation
(\ref{eq:max-receive-Q}), which measures the maximum amount that
can be received by network $Q$.
\end{enumerate}
\end{itemize}

\begin{itemize}
\item The Table in Figure (\ref{fig:Intermediate-State-Variables}) shows
the amounts to be transferred across the networks and other key output
which can help to understand how the system is functioning. The first
15 rows represent scenarios based on manually selected values chosen
to illustrate certain special cases. The rest of the rows are based
on values selected randomly. The scenarios in each row are the same
as those corresponding to the rows given in Figures (\ref{fig:External-Inputs-Systems};
\ref{fig:Transfer-Amounts-Primary}). The first row indicates the
steps in the algorithm to which each column corresponds to.
\item The columns in Figure (\ref{fig:Intermediate-State-Variables}) -
corresponding to the scenarios in each row - represent the following
information respectively:
\end{itemize}
\begin{enumerate}
\item \textbf{TBD+Current\_P }represents $currTotalPlusTBDAmount_{Pt}$
given in Equation (\ref{eq:Current-Plus-TBD-P}), which is the sum
of the current invested amount and the amount to be deployed - net
deposit or withdrawal - on network $P$.
\item \textbf{TBD+Current\_Q }represents $currTotalPlusTBDAmount_{Qt}$
given in Equation (\ref{eq:Current-Plus-TBD-Q}), which is the sum
of the current invested amount and the amount to be deployed - net
deposit or withdrawal - on network $Q$.
\item \textbf{outsideBand\_PQ\_D }is the comparison of the amounts outside
bands - $amountOutsideBand_{Pt}$ and $amountOutsideBand_{Qt}$ given
in Equation (\ref{eq:Alternate-PQ}) - on networks $P$ and $Q$.
\item \textbf{outsideBand\_QP\_D }is the comparison of the amounts outside
bands - $amountOutsideBand_{Pt}$ and $amountOutsideBand_{Qt}$ given
in Equation (\ref{eq:Alternate-QP}) - on networks $Q$ and $P$.
\item \textbf{outsideBand\_Positive\_P }indicates whether the amount outside
the lower and upper bands on network $P$ is positive or not (End-note
\ref{enu:Delta-value-of}).
\item \textbf{outsideBand\_Positive\_Q }indicates whether the amount outside
the lower and upper bands on network $Q$ is positive or not (End-note
\ref{enu:Delta-value-of}).
\item \textbf{transfer\_PQ\_First\_D} is the value of the first term in
Equation (\ref{eq:Alternate-PQ}).
\item \textbf{transfer\_PQ\_First} is the value of the first term in Equation
(\ref{eq:Transfer-Simple-PQ}).
\item \textbf{transfer\_PQ\_Second} is the value of the second term in Equation
(\ref{eq:Alternate-PQ}).
\item \textbf{transfer\_QP\_First\_D} is the value of the first term in
Equation (\ref{eq:Alternate-QP}).
\item \textbf{transfer\_QP\_First} is the value of the first term in Equation
(\ref{eq:Transfer-Simple-QP}).
\item \textbf{transfer\_QP\_Second} is the value of the second term in Equation
(\ref{eq:Alternate-QP}).
\item \textbf{Total-MinCapacity\_P }is the difference between the sum of
the current invested and to be deployed amounts - \textbf{TBD+Current\_P,}
$currTotalPlusTBDAmount_{Pt}$ - and the minimum capacity on network
$P$ - $minNetworkCapacity_{Pt}$ given in Equation (\ref{eq:minNetworkCapacity_P}).
\item \textbf{Total-MaxCapacity\_P }is the difference between the sum of
the current invested and to be deployed amounts - \textbf{TBD+Current\_P,}
$currTotalPlusTBDAmount_{Pt}$ - and the maximum capacity on network
$P$ - $maxNetworkCapacity_{Pt}$ given in Equation (\ref{eq:maxNetworkCapacity_P}).
\item \textbf{Total-MinCapacity\_Q }is the difference between the sum of
the current invested and to be deployed amounts - \textbf{TBD+Current\_Q,}
$currTotalPlusTBDAmount_{Qt}$ - and the minimum capacity on network
$Q$ - $minNetworkCapacity_{Qt}$ given in Equation (\ref{eq:minNetworkCapacity_Q}).
\item \textbf{Total-MaxCapacity\_Q }is the difference between the sum of
the current invested and to be deployed amounts - \textbf{TBD+Current\_Q,}
$currTotalPlusTBDAmount_{Qt}$ - and the maximum capacity on network
$Q$ - $maxNetworkCapacity_{Qt}$ given in Equation (\ref{eq:maxNetworkCapacity_Q}).
\end{enumerate}

\section{\label{sec:Conclusion}Implementation Pointers, Areas for Improvements
and Conclusions}

Wealth managers will select assets across multiple platforms such
that investors will get exposure to the whole suite of assets the
fund invests over all chains. Having positions on different chains
- and hence linking different networks - is one way of providing diversified
exposure to investors. The fund prices will have to be the same across
all the networks where the investment funds are deployed (Kashyap
2021-I). Maintaining investments across networks and adhering to certain
portfolio characteristics will require the transfer of funds between
networks using bridges.

There are several existing bridges that have been created by the platforms
themselves, but with limited ability in terms of which assets can
be transferred. Creation of new bridges for additional tokens are
being pursued by blockchain funds and third party providers. This
will be useful for the investors - community - and also act as a utility
for others. 

The investment machinery that does the fund management operations
- including interactions with investors by taking deposits and doing
redemptions - is best implemented on-chain as smart contracts (Wang
et al., 2018; Mohanta et al., 2018; Zou et al., 2019; Zheng et al.,
2020; End-note \ref{enu:A-smart-contract}). The bridge algorithm
can be a stand alone component - external to the blockchain system
- that reads the investment variables and outputs the amount to be
transferred, which will be used by fund personnel to authorize the
necessary bridge transactions. The weight calculation engine will
also be a separate component which interacts with the bridge calculation
routines and provides the relevant information to the portfolio teams.

We outline several areas for improving the basic ideas discussed here:
\begin{itemize}
\item The formulae we have developed will need to be modified when short
positions are to be allowed. The more general derivations in Kashyap
(2021-I) can serve as examples since they can handle shorting. The
extensions to handle shorting should be fairly easy, though several
conditions need to be verified thoroughly regarding weights and current
total amounts which can become negative.
\item Significant improvements can be made in forecasting methods as discussed
in End-note (\ref{enu:Numerous-alternate-Formulations}). 
\item Gas fees on individual networks have not been explicitly included
in our derivations since market participants can be expected to move
funds to lower cost networks over time. If TVL changes on networks
- to reflect gas fees - are not happening fast enough, we need to
consider formulae modifications. 
\item This mechanism can be extended to more than two networks. Sort the
networks based on their need to receive or send funds (Step \ref{enu:Transfer-Need}).
Starting with the network having the highest need in one direction,
match it with networks having the greatest transfer requirements in
the other direction. Continue network after network - pairwise round
robin fashion for a fixed number of iterations - till we satisfy network
requirements in one direction or exhaust bridge capacities. Once most
of the networks are saturated - in terms of fund transfer needs -
a few networks might still have pending amounts. These amounts can
be fulfilled by extending the capacity on some networks or waiting
for the next rebalancing event when the cycle of fund flows will commence
all over again.
\item For the sake of brevity, we have focused on the central elements of
our technique. The actual technical implementation will have to cover
several specialized scenarios, nuances or other constraints. Additional
checks pertaining to division by zero and other such cases need to
be considered in the software (End-note \ref{enu:Software-Testing-Validation}). 
\end{itemize}
A detailed algorithm has been developed to transfer assets when there
are network bridge constraints. The bridge utilization is dynamically
altered depending on external conditions. Such a requirement arises
naturally when attempting to provide risk managed wealth funds across
multiple blockchains. Several enhancements will be pursued in later
versions. Given the limitations of bridge technology, careful usage
of bridges is prudent with gradual increases in amounts transferred
depending on criticality of fund flow needs and technological enhancements. 

Clearly, there are external dependencies that become important: improvements
in bridge technology, the use of advanced strategies such as indices
or vaults - which will affect fund flows, security improvements or
other innovations than what is possible using bridges. When there
is money being moved around, there will be regulatory scrutiny and
watching out for upcoming policies will be something crucial in the
crypto-currency domain.

The pace of technological advancement is quite rapid in the blockchain
landscape. We have to revisit and review the environmental conditions
and our fund movement requirements constantly to ensure that portfolio
management goals are satisfied.

\section{\label{sec:Explanations-and-End-notes}Explanations and End-notes}
\begin{enumerate}
\item Acknowledgements and Clarifications: 
\begin{enumerate}
\begin{doublespace}
\item Numerous seminar participants, particularly at a few meetings of the
econometric society and various finance organizations, provided suggestions
to improve the material in this paper. 
\item The views and opinions expressed in this article, along with any mistakes,
are mine alone and do not necessarily reflect the official policy
or position of either of my affiliations or any other agency.
\end{doublespace}
\item Despite all the uncertainty in almost everything we do, we could surely
surmise that numerous others, (including members from the industry,
academia and elsewhere?), might have contributed intentionally and
/ or unintentionally to the creation of this piece. Their omission
from the acknowledgments is mostly unintentional and certainly unavoidable.
\end{enumerate}
\item \label{enu:Many-protocols-with}The invention of Bitcoin in 2008,
and the subsequent launch of the currency in 2009, is no doubt a landmark
event permanently etched in the history of technological innovations.
This seminal event is opening frontiers that are set to transform
all aspects of human interactions (Nakamoto 2008; Narayanan \& Clark
2017; Chen 2018; Monrat, et al. 2019). It has opened the floodgates
for innovations seeking to add different aspects of business and human
experiences onto the blockchain (Lindman et al., 2017; Kuo et al.,
2019; Lu 2019; Prewett et al., 2020; Zamani et al., 2020; Briola et
al., 2023). The rest, as they say, is history.
\begin{enumerate}
\item The following terms are important to understand how blockchain operates:
The Ledger; Linked Time stamping; Merkle Trees; Byzantine fault tolerance;
Proof Of Work.
\item A blockchain is a distributed ledger with growing lists of records
(blocks) that are securely linked together via cryptographic hashes.
Each block contains a cryptographic hash of the previous block, a
timestamp, and transaction data (generally represented as a Merkle
tree, where data nodes are represented by leaves). Since each block
contains information about the previous block, they effectively form
a chain (compare linked list data structure), with each additional
block linking to the ones before it. Consequently, blockchain transactions
are irreversible in that, once they are recorded, the data in any
given block cannot be altered retroactively without altering all subsequent
blocks. \href{https://en.wikipedia.org/wiki/Blockchain}{Blockchain,  Wikipedia Link}
\item In cryptography and computer science, a hash tree or Merkle tree is
a tree in which every \textquotedbl leaf\textquotedbl{} (node) is
labelled with the cryptographic hash of a data block, and every node
that is not a leaf (called a branch, inner node, or inode) is labelled
with the cryptographic hash of the labels of its child nodes. A hash
tree allows efficient and secure verification of the contents of a
large data structure. A hash tree is a generalization of a hash list
and a hash chain. \href{https://en.wikipedia.org/wiki/Merkle_tree}{Merkle Tree,  Wikipedia Link}
\item Although blockchain records are not unalterable, since blockchain
forks are possible, blockchains may be considered secure by design
and exemplify a distributed computing system with high Byzantine fault
tolerance. 
\item A Byzantine fault (also Byzantine generals problem, interactive consistency,
source congruency, error avalanche, Byzantine agreement problem, and
Byzantine failure) is a condition of a computer system, particularly
distributed computing systems, where components may fail and there
is imperfect information on whether a component has failed. The term
takes its name from an allegory, the \textquotedbl Byzantine generals
problem\textquotedbl , developed to describe a situation in which,
in order to avoid catastrophic failure of the system, the system's
actors must agree on a concerted strategy, but some of these actors
are unreliable. \href{https://en.wikipedia.org/wiki/Byzantine_fault}{Byzantine Fault,  Wikipedia Link}
\item Proof of work (PoW) is a form of cryptographic proof in which one
party (the prover) proves to others (the verifiers) that a certain
amount of a specific computational effort has been expended. Verifiers
can subsequently confirm this expenditure with minimal effort on their
part (Jakobsson \& Juels 1999). The purpose of proof-of-work algorithms
is not proving that certain work was carried out or that a computational
puzzle was \textquotedbl solved\textquotedbl , but deterring manipulation
of data by establishing large energy and hardware-control requirements
to be able to do so. \href{https://en.wikipedia.org/wiki/Proof_of_work}{Proof of Work,  Wikipedia Link}
\item Proof-of-work systems have been criticized by environmentalists for
their energy consumption. Several alternatives are being developed
due to the environment concerns of to PoW algorithms (Miraz et al.,
2021; Dimitri 2022). 
\item Proof-of-stake (PoS) protocols are a class of consensus mechanisms
for blockchains that work by selecting validators in proportion to
their quantity of holdings in the associated cryptocurrency (Saleh
2021; Wendl et al., 2023). This is done to avoid the computational
cost of proof-of-work (POW) schemes. The first functioning use of
PoS for cryptocurrency was Peer-coin in 2012, although the scheme,
on the surface, still resembled a POW. \href{https://en.wikipedia.org/wiki/Proof_of_stake}{Proof of Stake,  Wikipedia Link}
\item Many protocols with wonderful possibilities are being developed since
the creation of Bitcoin. At this time, ETH, BSC and Polygon are good
candidates to first launch investment funds (Caldarelli 2021; Donmez
\& Karaivanov 2022; Busayatananphon \& Boonchieng 2022; Urquhart 2022;
Connors \& Sarkar 2023). These three protocols are good candidates
for starting out given the remarkable progress they have made, the
stability they bring to this space and the similarity they offer in
terms of technological requirements. All three of them are EVM (Ethereum
Virtual Machine) compatible, making it relatively straightforward
to start using another of these platforms once a product is built
for one of these chains (Jia \& Yin 2022).That said: ETH with high
gas fees, BSC with some vulnerabilities in its choice of validators,
Polygon with scalability issues at times represent challenges that
are inherent in any technology saga. Numerous small tweaks and entire
redesigns of architectural frameworks are being undertaken with these
networks and their future looks promising.
\item Launching an investment product in phases is practical so that we
can thoroughly test on each platform and resolve any issues related
to each blockchain system. 
\item Solana, Fantom, Harmony One, Avalanche are some chains, which are
showing a lot of promise, and should feature actively in any plans
to deploy products and invest in assets on these platforms. Several
other platforms could also be on the immediate radar. As and when
promising investment opportunities arise on newer chains, it is prudent
to be prepared to monetize that.
\item \label{enu:CoinMarketCap}CoinMarketCap is a leading price-tracking
website for crypto-assets in the cryptocurrency space. Its mission
is to make crypto discoverable and efficient globally by empowering
retail users with unbiased, high quality and accurate information
for drawing their own informed conclusions. It was founded in May
2013 by Brandon Chez. \href{https://coinmarketcap.com/about/}{CoinMarketCap, Website Link}
\item \label{enu:Crypto-Ranking}A ranking of cryptocurrencies, including
symbols for the various tokens, by market capitalization is available
on the CoinMarketCap website. We are using the data as of May-25-2022,
when the first version of this article was written. \href{https://coinmarketcap.com}{CoinMarketCap Cryptocurrency Ranking,  Website Link}
\end{enumerate}
\item \label{enu:Excessive-financialization--}Excessive financialization
- and market risks causing financial instability (Acharya \& Richardson
2009; Reinhart \& Rogoff 2009; Bonizzi 2013; Palley 2013; Aalbers
2015; Davis \& Kim 2015) - can also be measured by comparing amounts
being transferred across networks and the amounts invested in those
networks. 
\begin{enumerate}
\item Hedge Funds and Mutual Funds form a core component of the traditional
financial system and hence monitoring their operations could indicate
excessive financialization. Replicating some features of Hedge Funds
and Mutual Funds on blockchain would be crucial to ensure properly
functioning decentralized wealth management platforms (Cai 2018; Peterson
2018; Arshadi 2019; Schär 2021; Kashyap 2021-I; 2021-II; Dos Santos
et al., 2022; Agarwal et al., 2009; Stulz 2007).
\end{enumerate}
\item \label{enu:TVL}In decentralized finance, Total value locked represents
the number of assets that are currently being staked in a specific
protocol.\href{https://coinmarketcap.com/alexandria/glossary/total-value-locked-tvl}{Total Value Locked, CoinMarketCap Link} 
\begin{enumerate}
\item For a blockchain investment fund this would be the total investment
funds received or the total amount of money being managed by the fund.
\item \label{enu:Finance-AUM}In finance, assets under management (AUM),
sometimes called fund under management, measures the total market
value of all the financial assets which an individual or financial
institution—such as a mutual fund, venture capital firm, or depository
institution—or a decentralized network protocol controls, typically
on behalf of a client. \href{https://en.wikipedia.org/wiki/Assets_under_management}{Assets Under Management, Wikipedia Link}
\end{enumerate}
\item \label{enu:Blockchain-Bridges}Blockchain bridges work just like the
bridges we know in the physical world. Just as a physical bridge connects
two physical locations, a blockchain bridge connects two blockchain
ecosystems. Bridges facilitate communication between blockchains through
the transfer of information and assets (Belchior et al., 2021; Qasse
et al., 2019; Schulte et al., 2019; Hafid et al., 2020; Zhou et al.,
2020; Stone 2021; Li et al., 2022). \href{https://ethereum.org/en/bridges/}{Blockchain Bridges,  Ethereum.Org Website Link};
\href{https://hacken.io/discover/blockchain-bridges/}{Blockchain Bridges 101,  hacken.io Website Link}
\begin{itemize}
\item The following factors are immediately applicable when attempting a
bridge transfer: 1) Time Taken for Transfer to Complete 2) Using Multiple
Wallets, and 3) Exit Liquidity. 
\item During times of network congestion, higher times might needed to complete
the transfer (Sokolov 2021; Dotan et al., 2021; Jiang et al., 2022).
Latency is the time taken for data to reach from one chain to another.
\href{[https://ethereum.org/en/bridges/\%7C\%7CBlockchain\%20Bridges,\%20\%20Ethereum.Org\%20Website\%20Link]}{Blockchain Bridges Introduction,  Medium Website Link};
\href{https://medium.com/1kxnetwork/blockchain-bridges-5db6afac44f8}{Blockchain Bridges Networks,  Medium Website Link}.
\item A cryptocurrency wallet is a device, physical medium, program or a
service which stores the public and/or private keys for cryptocurrency
transactions. In addition to this basic function of storing the keys,
a cryptocurrency wallet more often offers the functionality of encrypting
and/or signing information (Suratkar et al., 2020). \href{https://en.wikipedia.org/wiki/Cryptocurrency_wallet}{Cryptocurrency Wallet,  Wikipedia Link} 
\item Exit Liquidity refers to the amount of tokens available on the destination
network. The amount that can be transferred is limited by the exit
liquidity. The bridge capacity - tokens that can be sent or received
- at any time depends on both the amount of tokens at the sending
and receiving platform. Even if the sending platform has token availability,
unless the receiving platform has sufficient liquidity the bridge
transaction will not complete and can even fail. Sometimes, when sufficient
liquidity is not there, placeholder tokens are given which can be
converted to the original tokens when liquidity is replenished.
\end{itemize}
\item \label{enu:Decentralized-finance}Decentralized finance (often stylized
as DeFi) offers financial instruments without relying on intermediaries
such as brokerages, exchanges, or banks by using smart contracts on
a blockchain. \href{https://en.wikipedia.org/wiki/Decentralized_finance}{Decentralized Finance (DeFi), Wikipedia Link}
\begin{enumerate}
\item \label{enu:Centralized-cryptocurrency-excha}Centralized cryptocurrency
exchanges (CEX), or just centralized exchanges, act as an intermediary
between a buyer and a seller and make money through commissions and
transaction fees. You can imagine a CEX to be similar to a stock exchange
but for digital assets. \href{https://corporatefinanceinstitute.com/resources/cryptocurrency/cryptocurrency-exchanges/}{Centralized Cryptocurrency Exchanges (CEX),  Wikipedia Link};
\href{https://www.investopedia.com/tech/what-are-centralized-cryptocurrency-exchanges/}{Centralized Cryptocurrency Exchanges (CEX),  Investopedia Link};
\href{https://www.coindesk.com/learn/what-is-a-cex-centralized-exchanges-explained/}{Centralized Cryptocurrency Exchanges (CEX),  CoinDesk Link};
\href{https://www.coindesk.com/learn/centralized-exchange-cex-vs-decentralized-exchange-dex-whats-the-difference/}{CEX vs DEX Difference,  CoinDesk Link}
\item \label{enu:Decentralized-exchanges-(DEX)}Decentralized exchanges
(DEX) are a type of cryptocurrency exchange which allows for direct
peer-to-peer cryptocurrency transactions to take place without the
need for an intermediary. \href{https://en.wikipedia.org/wiki/Decentralized_finance\#Decentralized_exchanges}{Decentralized Exchanges (DEX),  Wikipedia Link};
\href{https://www.coindesk.com/learn/what-is-a-dex-how-decentralized-crypto-exchanges-work/}{Decentralized Cryptocurrency Exchanges (DEX),  CoinDesk Link}
\item \label{enu:Types-Yield-Enhancement-Services}The following are the
four main types of blockchain yield enhancement services. We can also
consider them as the main types of financial products available in
decentralized finance: 
\begin{enumerate}
\item \label{enu:Single-sided-staking-allows}Single-Sided Staking: This
allows users to earn yield by providing liquidity for one type of
asset, in contrast to liquidity provisioning on AMMs, which requires
a pair of assets. \href{https://docs.saucerswap.finance/features/single-sided-staking}{Single Sided Staking,  SuacerSwap Link}
\begin{enumerate}
\item Bancor is an example of a provider who supports single sided staking.
Bancor natively supports Single-Sided Liquidity Provision of tokens
in a liquidity pool. This is one of the main benefits to liquidity
providers that distinguishes Bancor from other DeFi staking protocols.
Typical AMM liquidity pools require a liquidity provider to provide
two assets. Meaning, if you wish to deposit \textquotedbl TKN1\textquotedbl{}
into a pool, you would be forced to sell 50\% of that token and trade
it for \textquotedbl TKN2\textquotedbl . When providing liquidity,
your deposit is composed of both TKN1 and TKN2 in the pool. Bancor
Single-Side Staking changes this and enables liquidity providers to:
Provide only the token they hold (TKN1 from the example above) Collect
liquidity providers fees in TKN1. \href{https://docs.bancor.network/about-bancor-network/faqs/single-side-liquidity}{Single Sided Staking,  Bancor Link}
\end{enumerate}
\item \label{enu:AMM-Liquidity-Pairs}AMM Liquidity Pairs (AMM LP): A constant-function
market maker (CFMM) is a market maker with the property that that
the amount of any asset held in its inventory is completely described
by a well-defined function of the amounts of the other assets in its
inventory (Hanson 2007). \href{https://en.wikipedia.org/wiki/Constant_function_market_maker}{Constant Function Market Maker,  Wikipedia Link}

This is the most common type of market maker liquidity pool. Other
types of market makers are discussed in Mohan (2022). All of them
can be grouped under the category Automated Market Makers. Hence the
name AMM Liquidity Pairs. A more general discussion of AMMs, without
being restricted only to the blockchain environment, is given in (Slamka,
Skiera \& Spann 2012).
\item \label{enu:LP-Token-Staking:}LP Token Staking: LP staking is a valuable
way to incentivize token holders to provide liquidity. When a token
holder provides liquidity as mentioned earlier in Point (\ref{enu:AMM-Liquidity-Pairs})
they receive LP tokens. LP staking allows the liquidity providers
to stake their LP tokens and receive project tokens tokens as rewards.
This mitigates the risk of impermanent loss and compensates for the
loss. \href{https://defactor.com/liquidity-provider-staking-introduction-guide/}{Liquidity Provider Staking,  DeFactor Link}
\begin{enumerate}
\item Note that this is also a type of single sided staking discussed in
Point (\ref{enu:Single-sided-staking-allows}). The key point to remember
is that the LP Tokens can be considered as receipts for the crypto
assets deposits in an AMM LP Point (\ref{enu:AMM-Liquidity-Pairs}).
These LP Token receipts can be further staked to generate additional
yield.
\end{enumerate}
\item \label{enu:Lending:-Crypto-lending}Lending: Crypto lending is the
process of depositing cryptocurrency that is lent out to borrowers
in return for regular interest payments. Payments are typically made
in the form of the cryptocurrency that is deposited and can be compounded
on a daily, weekly, or monthly basis. \href{https://www.investopedia.com/crypto-lending-5443191}{Crypto Lending,  Investopedia Link};
\href{https://defiprime.com/decentralized-lending}{DeFi Lending,  DeFiPrime Link};
\href{https://crypto.com/price/categories/lending}{Top Lending Coins by Market Capitalization,  Crypto.com Link}
\begin{enumerate}
\item Crypto lending is very common on decentralized finance projects and
also in centralized exchanges. Centralized cryptocurrency exchanges
are online platforms used to buy and sell cryptocurrencies. They are
the most common means that investors use to buy and sell cryptocurrency
holdings. \href{https://www.investopedia.com/tech/what-are-centralized-cryptocurrency-exchanges/}{Centralized Cryptocurrency Exchanges,  Investopedia Link}
\item Lending is a very active area of research both on blockchain and off
chain (traditional finance) as well (Cai 2018; Zeng et al., 2019;
Bartoletti, Chiang \& Lafuente 2021; Gonzalez 2020; Hassija et al.,
2020; Patel et al. , 2020). 
\end{enumerate}
\end{enumerate}
\item Investment strategies and Funds flows on DeFi have to deal with additional
constraints - compared to traditional finance - such as bridge limitations,
asset availability on individual networks and gas fees (Bender et
al., 2010; Bass et al., 2017; Liu 2019; Monrat et al., 2019; Zarir
et al., 2021; Bertsimas \& Lo 1998; Almgren \& Chriss 2001; Fung et
al., 2022). Asset allocation techniques developed for mainstream financial
portfolios - (Donohue \& Yip 2003; Tokat \& Wicas 2007; Calvet et
al., 2009) - have to be tailored for blockchain nuances.
\end{enumerate}
\item \label{enu:Networks-Rebalancing}The issue with networks that do not
get rebalanced often is that the amount collected on that network
might not get transferred and deployed on other networks that get
rebalanced more often. Similarly, for withdrawals the wait can be
longer on networks that have longer time periods between rebalancing
events. We can choose to just transfer assets without rebalancing,
but that can produce some issues in terms of portfolio risk management
and fund flow requirements.
\begin{enumerate}
\item Also, there will be different degrees of correlation between prices
across different chains depending on the extent of inter-connectedness
between them. As the fund flow increases across existing chains, it
is highly likely that the movements will increase in lock steps. The
greater overlap between chains in terms of asset movements will also
bring about the risk for a drastic drop in total value invested on
any chain, if that particular network starts to lose trust and get
abandoned. Initially, frictions that will impede fund movements will
serve the best interest of certain parties, such as network owners
who want more funds locked on their platforms. But as competitive
pressures erode the frictions, they will later exacerbate certain
other risks - such as the possibility of entire networks losing funds
in a short period of time.
\end{enumerate}
\item \label{enu:Front-running,-also}Front running, also known as tailgating,
is the practice of entering into an equity (stock) trade, option,
futures contract, derivative, or security-based swap to capitalize
on advance, nonpublic knowledge of a large (\textquotedbl block\textquotedbl )
pending transaction that will influence the price of the underlying
security (Bernhardt \& Taub 2008; Baum et al., 2022). \href{https://en.wikipedia.org/wiki/Front_running}{Front Running,  Wikipedia Link}
\item \label{enu:Numerous-alternate-Formulations}Clearly, numerous alternate
formulations can consider varying probability distributions and other
such complexities (Fernandez 1981; Hamilton 2020). But we give preference
to simple mechanisms that bring robustness and to ensure that the
system operates well under a variety of conditions. 
\begin{enumerate}
\item The best estimator for any system is the system itself and hence using
the historical data directly - without having to estimate numerous
parameters and then forecast values wherein information can be lost
- should be the preferred method (Kashyap 2022-II). 
\item When using historical data, we need to make sure that we use information
after the system has reached a somewhat stable phase after a few months
of operation. Also the forecast time period should be much smaller
than the time period over which historical data is used.
\item We wish to point out here a key difference between science and engineering.
Science is about understanding existing systems. Engineering is about
building new systems - based on knowledge of other systems, that is
using science - that can operate effectively under different scenarios.
Hence, simplicity which leads to robustness is recommended. We have
also tried to ensure that the system will govern itself over time
based on the evolution of various metrics with the least amount of
external data dependencies.
\item The range based models discussed in Kashyap (2022) are based on a
wider set of techniques termed: Randoptimization. The limitations
of optimization methodologies and the need for range based methodologies
- which introduce randomness in the decision process - are discussed
in detail in the series: ``Fighting Uncertainty with Uncertainty''
Kashyap (2016). The minimum and maximum asset weights we have discussed
in the main text are based on this idea of operating a system within
a range as opposed to pinning down operational parameters to a single
value. The range of values is prudent to use due to the errors that
exist around the estimates we obtain for an ideal value. Clearly,
the weight range we can use to mitigate network constraints is dependent
upon the estimation errors in the corresponding weight optimization
process. Conversely, depending on the extent of the constraints for
funds flows in a network, we can decide the width of the range we
can tolerate for the weights.
\item We model variables that only take positive values as Geometric Brownian
Motions (GBMs). The uncertainty in these variables is introduced by
estimating the corresponding parameters of the GBM. The parameters
can also be sampled from suitable log normal distributions or by sampling
from suitable absolute normal distributions with their own parameters
(Equations: \ref{eq:GBM}; \ref{eq:Absolute-Normal-Distribution}). 
\item Norstad (1999) has a technical discussion of the normal and log normal
distributions. Hull \& Basu (2016) provide an excellent account of
using GBMs to model stock prices and other time series that are always
positive. It is worth noting that the starting value, mean and standard
deviation of the time series can themselves be simulations from other
appropriately chosen uniform distributions. Some of the above variables
can be modeled as Poisson distributions or we can simply consider
them as the absolute value of a normal distribution with appropriately
chosen units.
\item In statistics, a normal distribution or Gaussian distribution is a
type of continuous probability distribution for a real-valued random
variable. The general form of its probability density function is

\begin{equation}
{\displaystyle f(x)={\frac{1}{\sigma{\sqrt{2\pi}}}}e^{-{\frac{1}{2}}\left({\frac{x-\mu}{\sigma}}\right)^{2}}}
\end{equation}
 The parameter $\mu$ is the mean or expectation of the distribution
(and also its median and mode), while the parameter $\sigma$ is its
standard deviation. \href{https://en.wikipedia.org/wiki/Normal_distribution}{Normal Distribution,  Wikipedia Link}
\item In probability theory and statistics, the Poisson distribution is
a discrete probability distribution that expresses the probability
of a given number of events occurring in a fixed interval of time
or space if these events occur with a known constant mean rate and
independently of the time since the last event. \href{https://en.wikipedia.org/wiki/Poisson_distribution}{Poisson Distribution,  Wikipedia Link}

A discrete random variable $X$ is said to have a Poisson distribution,
with parameter ${\displaystyle \lambda>0,}$ if it has a probability
mass function given by:

\begin{equation}
{\displaystyle f(k;\lambda)=\Pr(X{=}k)={\frac{\lambda^{k}e^{-\lambda}}{k!}},}
\end{equation}
where k is the number of occurrences (${\displaystyle k=0,1,2,\ldots}$).
$e$ is Euler's number ($e=2.71828\ldots$). $!$ is the factorial
function.
\item A geometric Brownian motion (GBM) (also known as exponential Brownian
motion) is a continuous-time stochastic process in which the logarithm
of the randomly varying quantity follows a Brownian motion (also called
a Wiener process) with drift. \href{https://en.wikipedia.org/wiki/Geometric_Brownian_motion}{Geometric Brownian Motion,  Wikipedia Link}
\item A GBM is characterized as below. $S_{it}$ is the stochastic process
that follows a GBM by satisfying the below stochastic differential
equation (Equation: \ref{eq:GBM}). $S_{i}$ could be the price -
or another variable that always takes positive values - of the $i^{th}$
security. $\mu_{S_{i}}$ is the drift and $\sigma_{S_{i}}$ is the
volatility. $W_{t}^{S_{i}}$ is the Weiner Process governing the $S_{i}^{th}$
variable. 

\begin{doublespace}
\begin{equation}
\text{Geometric Brownian Motion }\equiv\frac{dS_{it}}{S_{it}}=\mu_{S_{i}}dt+\sigma_{S_{i}}dW_{t}^{S_{i}}\label{eq:GBM}
\end{equation}
Alternately, we could sample the variable values, $S_{it}$, from
an absolute normal distribution with mean, $\mu_{S_{i}}$, and variance,
$\sigma_{S_{i}}^{2}$, as shown in Equation (\ref{eq:Absolute-Normal-Distribution}).
The folded normal distribution is a probability distribution related
to the normal distribution. Given a normally distributed random variable
$X$ with mean $\mu$ and variance $\sigma^{2}$, the random variable
$Y=|X|$ has a folded normal distribution. The distribution is called
\textquotedbl folded\textquotedbl{} because probability mass to the
left of $x=0$ is folded over by taking the absolute value. \href{https://en.wikipedia.org/wiki/Folded_normal_distribution}{Folded Normal Distribution,  Wikipedia Link}

\begin{align}
\text{Alternately},\;S_{it} & \sim\left|N\left(\mu_{S_{i}},\sigma_{S_{i}}^{2}\right)\right|,\;\text{Absolute\;\ Normal\;\ Distribution}\label{eq:Absolute-Normal-Distribution}
\end{align}
The simulation seeds - the parameters - are chosen so that the drift
and volatility we get for the variables are similar to what would
be observed in practice. 
\end{doublespace}
\end{enumerate}
\item \label{enu:Delta-value-of}The value of $\triangle=0.0001$ should
suffice. This means that we need to ensure that the system will not
allow withdraw requests or other amounts smaller than $\triangle=0.0001$
such that $\left|amountOutsideBand_{Pt,Qt}\right|\geq\triangle$.
By using the alternate formulations (Equations: \ref{eq:Alternate-PQ};
\ref{eq:Alternate-QP}) if the withdraw on a network is larger than
the current amount on that network and also larger than the amount
above the maximum band on the other network, we transfer up-to the
minimum band from the other network to the network needing the extra
to satisfy the withdraw requests. Note that, 
\begin{equation}
\left[\frac{\max\left(amountOutsideBand_{Pt}+\triangle,0\right)}{\left|amountOutsideBand_{Pt}\right|+\triangle}\right]=\begin{cases}
1, & \text{if, }amountOutsideBand_{Pt}\geq0\\
0, & \text{if, }amountOutsideBand_{Pt}<0
\end{cases}
\end{equation}
\item \label{enu:A-smart-contract}A smart contract is a computer program
or a transaction protocol that is intended to automatically execute,
control or document events and actions according to the terms of a
contract or an agreement. The objectives of smart contracts are the
reduction of need for trusted intermediators, arbitration costs, and
fraud losses, as well as the reduction of malicious and accidental
exceptions (Wang et al., 2018; Mohanta et al., 2018; Zou et al., 2019;
Zheng et al., 2020). \href{https://en.wikipedia.org/wiki/Smart_contract}{Smart Contract,  Wikipedia Link}
\item \label{enu:Software-Testing-Validation}We would like to highlight
the following points to help with the actual coding of the software
(Boehm 1983; Balci 1995; Denning 2005; Desikan \& Ramesh 2006; Sargent
2010; Green \& Ledgard 2011; Knuth 2014). The algorithm we have provided
acts mostly as detailed implementation guidelines. Many cases and
error conditions need to be handled appropriately during implementation.
Alternate implementation simplifications, time conventions, and counters
are possible and can be accommodated accordingly. There might even
be some issues - or bugs - with the variables, counters and timing.
These are due to limitations of not actually testing scenarios using
a full fledged software system. But the gist of what we have provided
should carry over to the coding stage with very little changes. Conditional
statements such as - if ... then ... else - can be used depending
on the implementation language and other efficiency considerations
as necessary.
\item \label{enu:Uniform-Distribution}In probability theory and statistics,
the continuous uniform distributions or rectangular distributions
are a family of symmetric probability distributions. Such a distribution
describes an experiment where there is an arbitrary outcome that lies
between certain bounds. The bounds are defined by the parameters,
$a$ and $b$, which are the minimum and maximum values (Dekking et
al., 2005). The interval can either be closed (i.e. $[a,b]$) or open
(i.e. $(a,b)$). Therefore, the distribution is often abbreviated
${\displaystyle U(a,b),}$ where $U$ stands for uniform distribution
(Walpole et al., 1993).  \href{https://en.wikipedia.org/wiki/Continuous_uniform_distribution}{Continuous Uniform Distribution,  Wikipedia Link}
\end{enumerate}

\section{\label{sec:References}References}
\begin{itemize}
\item Aalbers, M. B. (2015). The potential for financialization. Dialogues
in Human Geography, 5(2), 214-219.
\item Acharya, V. V., \& Richardson, M. P. (Eds.). (2009). Restoring financial
stability: how to repair a failed system (Vol. 542). John Wiley \&
Sons.
\item Agarwal, V., Boyson, N. M., \& Naik, N. Y. (2009). Hedge funds for
retail investors? An examination of hedged mutual funds. Journal of
Financial and Quantitative Analysis, 44(2), 273-305.
\item Almgren, R., \& Chriss, N. (2001). Optimal execution of portfolio
transactions. Journal of Risk, 3, 5-40.
\item Arshadi, N. (2019). Application of Blockchain Protocol to Wealth Management.
The Journal of Wealth Management, 21(4), 122-129.
\item Balci, O. (1995, December). Principles and techniques of simulation
validation, verification, and testing. In Proceedings of the 27th
conference on Winter simulation (pp. 147-154).
\item Bartoletti, M., Chiang, J. H. Y., \& Lafuente, A. L. (2021). SoK:
lending pools in decentralized finance. In Financial Cryptography
and Data Security. FC 2021 International Workshops: CoDecFin, DeFi,
VOTING, and WTSC, Virtual Event, March 5, 2021, Revised Selected Papers
25 (pp. 553-578). Springer Berlin Heidelberg.
\item Bass, R., Gladstone, S., \& Ang, A. (2017). Total portfolio factor,
not just asset, allocation. The Journal of Portfolio Management, 43(5),
38-53.
\item Baum, C., Chiang, J. H. Y., David, B., Frederiksen, T. K., \& Gentile,
L. (2021). Sok: Mitigation of front-running in decentralized finance.
Cryptology ePrint Archive.
\item Belchior, R., Vasconcelos, A., Guerreiro, S., \& Correia, M. (2021).
A survey on blockchain interoperability: Past, present, and future
trends. ACM Computing Surveys (CSUR), 54(8), 1-41.
\item Bender, J., Briand, R., Nielsen, F., \& Stefek, D. (2010). Portfolio
of risk premia: A new approach to diversification. The Journal of
Portfolio Management, 36(2), 17-25.
\item Bernhardt, D., \& Taub, B. (2008). Front-running dynamics. Journal
of Economic Theory, 138(1), 288-296.
\begin{doublespace}
\item Bertsimas, D., \& Lo, A. W. (1998). Optimal control of execution costs.
Journal of Financial Markets, 1(1), 1-50.
\end{doublespace}
\item Boehm, B. W. (1983). Seven basic principles of software engineering.
Journal of Systems and Software, 3(1), 3-24.
\item Bonizzi, B. (2013). Financialization in developing and emerging countries:
a survey. International journal of political economy, 42(4), 83-107.
\item Briola, A., Vidal-Tomás, D., Wang, Y., \& Aste, T. (2023). Anatomy
of a Stablecoin’s failure: The Terra-Luna case. Finance Research Letters,
51, 103358.
\item Busayatananphon, C., \& Boonchieng, E. (2022, January). Financial
technology DeFi protocol: A review. In 2022 Joint International Conference
on Digital Arts, Media and Technology with ECTI Northern Section Conference
on Electrical, Electronics, Computer and Telecommunications Engineering
(ECTI DAMT \& NCON) (pp. 267-272). IEEE.
\item Cai, C. W. (2018). Disruption of financial intermediation by FinTech:
a review on crowdfunding and blockchain. Accounting \& Finance, 58(4),
965-992.
\item Caldarelli, G. (2021). Wrapping trust for interoperability: A preliminary
study of wrapped tokens. Information, 13(1), 6.
\item Calvet, L. E., Campbell, J. Y., \& Sodini, P. (2009). Fight or flight?
Portfolio rebalancing by individual investors. The Quarterly journal
of economics, 124(1), 301-348.
\item Chen, Y. (2018). Blockchain tokens and the potential democratization
of entrepreneurship and innovation. Business horizons, 61(4), 567-575.
\item Connors, C., \& Sarkar, D. (2023). Survey of prominent blockchain
development platforms. Journal of Network and Computer Applications,
103650.
\item Davis, G. F., \& Kim, S. (2015). Financialization of the Economy.
Annual Review of Sociology, 41, 203-221.
\item Dekking, F. M., Kraaikamp, C., Lopuhaä, H. P., \& Meester, L. E. (2005).
A Modern Introduction to Probability and Statistics: Understanding
why and how (Vol. 488). London: springer.
\item Denning, P. J. (2005). Is computer science science?. Communications
of the ACM, 48(4), 27-31.
\item Desikan, S., \& Ramesh, G. (2006). Software testing: principles and
practice. Pearson Education India.
\item Dimitri, N. (2022). Consensus: Proof of Work, Proof of Stake and structural
alternatives. Enabling the Internet of Value: How Blockchain Connects
Global Businesses, 29-36.
\item Donmez, A., \& Karaivanov, A. (2022). Transaction fee economics in
the Ethereum blockchain. Economic Inquiry, 60(1), 265-292.
\item Donohue, C., \& Yip, K. (2003). Optimal portfolio rebalancing with
transaction costs. Journal of Portfolio Management, 29(4), 49.
\item Dos Santos, S., Singh, J., Thulasiram, R. K., Kamali, S., Sirico,
L., \& Loud, L. (2022, June). A new era of blockchain-powered decentralized
finance (DeFi)-a review. In 2022 IEEE 46th Annual Computers, Software,
and Applications Conference (COMPSAC) (pp. 1286-1292). IEEE.
\item Dotan, M., Pignolet, Y. A., Schmid, S., Tochner, S., \& Zohar, A.
(2021). Survey on blockchain networking: Context, state-of-the-art,
challenges. ACM Computing Surveys (CSUR), 54(5), 1-34.
\item Fernandez, R. B. (1981). A methodological note on the estimation of
time series. The Review of Economics and Statistics, 63(3), 471-476.
\item Fung, K., Jeong, J., \& Pereira, J. (2022). More to cryptos than bitcoin:
A GARCH modelling of heterogeneous cryptocurrencies. Finance research
letters, 47, 102544.
\item Gonzalez, L. (2020). Blockchain, herding and trust in peer-to-peer
lending. Managerial Finance, 46(6), 815-831.
\item Green, R., \& Ledgard, H. (2011). Coding guidelines: Finding the art
in the science. Communications of the ACM, 54(12), 57-63.
\item Hafid, A., Hafid, A. S., \& Samih, M. (2020). Scaling blockchains:
A comprehensive survey. IEEE access, 8, 125244-125262.
\item Hamilton, J. D. (2020). Time series analysis. Princeton university
press.
\item Hassija, V., Bansal, G., Chamola, V., Kumar, N., \& Guizani, M. (2020).
Secure lending: Blockchain and prospect theory-based decentralized
credit scoring model. IEEE Transactions on Network Science and Engineering,
7(4), 2566-2575.
\item Hanson, R. (2007). Logarithmic markets coring rules for modular combinatorial
information aggregation. The Journal of Prediction Markets, 1(1),
3-15.
\item Hull, J. C., \& Basu, S. (2016). Options, futures, and other derivatives.
Pearson Education India.
\item Jakobsson, M., \& Juels, A. (1999, September). Proofs of work and
bread pudding protocols. In Secure Information Networks: Communications
and Multimedia Security IFIP TC6/TC11 Joint Working Conference on
Communications and Multimedia Security (CMS’99) September 20–21, 1999,
Leuven, Belgium (pp. 258-272). Boston, MA: Springer US.
\item Jia, R., \& Yin, S. (2022, November). To EVM or Not to EVM: Blockchain
Compatibility and Network Effects. In Proceedings of the 2022 ACM
CCS Workshop on Decentralized Finance and Security (pp. 23-29).
\item Jiang, S., Li, Y., Wang, S., \& Zhao, L. (2022). Blockchain competition:
The tradeoff between platform stability and efficiency. European Journal
of Operational Research, 296(3), 1084-1097.
\item Kashyap, R. (2016). Fighting Uncertainty with Uncertainty. Available
at SSRN 2715424.
\item Kashyap, R. (2021-I). A Tale of Two Currencies: Cash and Crypto. Working
Paper.
\item Kashyap, R. (2021-II). Hedged Mutual Fund Blockchain Protocol: High
Water Marks During Low Market Prices. Working Paper.
\item Kashyap, R. (2022). Bringing Risk Parity To The Defi Party: A Complete
Solution To The Crypto Asset Management Conundrum. Initial Draft.
\item Kashyap, R. (2022-II). Options as Silver Bullets: Valuation of Term
Loans, Inventory Management, Emissions Trading and Insurance Risk
Mitigation using Option Theory. Annals of Operations Research, 315(2),
1175-1215.
\item Knuth, D. E. (2014). Art of computer programming, volume 2: Seminumerical
algorithms. Addison-Wesley Professional.
\item Kuo, T. T., Zavaleta Rojas, H., \& Ohno-Machado, L. (2019). Comparison
of blockchain platforms: a systematic review and healthcare examples.
Journal of the American Medical Informatics Association, 26(5), 462-478.
\item Lee, S. S., Murashkin, A., Derka, M., \& Gorzny, J. (2022). SoK: Not
Quite Water Under the Bridge: Review of Cross-Chain Bridge Hacks.
arXiv preprint arXiv:2210.16209.
\item Lindman, J., Tuunainen, V. K., \& Rossi, M. (2017). Opportunities
and risks of Blockchain Technologies–a research agenda.
\item Li, Y., Liu, H., \& Tan, Y. (2022, May). POLYBRIDGE: A Crosschain
Bridge for Heterogeneous Blockchains. In 2022 IEEE International Conference
on Blockchain and Cryptocurrency (ICBC) (pp. 1-2). IEEE.
\item Li, X., Jiang, P., Chen, T., Luo, X., \& Wen, Q. (2020). A survey
on the security of blockchain systems. Future generation computer
systems, 107, 841-853.
\item Liu, W. (2019). Portfolio diversification across cryptocurrencies.
Finance Research Letters, 29, 200-205.
\item Lu, Y. (2019). The blockchain: State-of-the-art and research challenges.
Journal of Industrial Information Integration, 15, 80-90.
\item Miraz, M. H., Excell, P. S., \& Rafiq, M. K. S. B. (2021). Evaluation
of green alternatives for blockchain proof-of-work (PoW) approach.
Annals of Emerging Technologies in Computing (AETiC), 54-59.
\item Mohan, V. (2022). Automated market makers and decentralized exchanges:
a DeFi primer. Financial Innovation, 8(1), 20.
\item Mohanta, B. K., Panda, S. S., \& Jena, D. (2018, July). An overview
of smart contract and use cases in blockchain technology. In 2018
9th international conference on computing, communication and networking
technologies (ICCCNT) (pp. 1-4). IEEE.
\item Monrat, A. A., Schelén, O., \& Andersson, K. (2019). A survey of blockchain
from the perspectives of applications, challenges, and opportunities.
IEEE Access, 7, 117134-117151.
\item Nakamoto, S. (2008). Bitcoin: A peer-to-peer electronic cash system.
Decentralized Business Review, 21260. 
\item Narayanan, A., \& Clark, J. (2017). Bitcoin’s academic pedigree. Communications
of the ACM, 60(12), 36-45.
\item Norstad, J. (1999). The normal and lognormal distributions.
\item Patel, S. B., Bhattacharya, P., Tanwar, S., \& Kumar, N. (2020). Kirti:
A blockchain-based credit recommender system for financial institutions.
IEEE Transactions on Network Science and Engineering, 8(2), 1044-1054.
\item Palley, T. I. (2013). Financialization: what it is and why it matters
(pp. 17-40). Palgrave Macmillan UK.
\item Peterson, M. (2018). Blockchain and the future of financial services.
The Journal of Wealth Management, 21(1), 124-131.
\item Prewett, K. W., Prescott, G. L., \& Phillips, K. (2020). Blockchain
adoption is inevitable—Barriers and risks remain. Journal of Corporate
accounting \& finance, 31(2), 21-28.
\item Qasse, I. A., Abu Talib, M., \& Nasir, Q. (2019, March). Inter blockchain
communication: A survey. In Proceedings of the ArabWIC 6th Annual
International Conference Research Track (pp. 1-6).
\item Reinhart, C. M., \& Rogoff, K. S. (2009). This time is different.
In This Time Is Different. princeton university press.
\item Saleh, F. (2021). Blockchain without waste: Proof-of-stake. The Review
of financial studies, 34(3), 1156-1190.
\item Sargent, R. G. (2010, December). Verification and validation of simulation
models. In Proceedings of the 2010 winter simulation conference (pp.
166-183). IEEE.
\item Schär, F. (2021). Decentralized finance: On blockchain-and smart contract-based
financial markets. FRB of St. Louis Review.
\item Scharfman, J. (2023). Decentralized finance (defi) fraud and hacks:
Part 2. In The Cryptocurrency and Digital Asset Fraud Casebook (pp.
97-110). Cham: Springer International Publishing.
\item Schulte, S., Sigwart, M., Frauenthaler, P., \& Borkowski, M. (2019).
Towards blockchain interoperability. In Business Process Management:
Blockchain and Central and Eastern Europe Forum: BPM 2019 Blockchain
and CEE Forum, Vienna, Austria, September 1–6, 2019, Proceedings 17
(pp. 3-10). Springer International Publishing.
\item Slamka, C., Skiera, B., \& Spann, M. (2012). Prediction market performance
and market liquidity: A comparison of automated market makers. IEEE
Transactions on Engineering Management, 60(1), 169-185.
\item Sokolov, K. (2021). Ransomware activity and blockchain congestion.
Journal of Financial Economics, 141(2), 771-782.
\item Stone, D. (2021). Trustless, privacy-preserving blockchain bridges.
arXiv preprint arXiv:2102.04660.
\item Stulz, R. M. (2007). Hedge funds: Past, present, and future. Journal
of Economic Perspectives, 21(2), 175-194.
\item Suratkar, S., Shirole, M., \& Bhirud, S. (2020, September). Cryptocurrency
wallet: A review. In 2020 4th international conference on computer,
communication and signal processing (ICCCSP) (pp. 1-7). IEEE.
\item Tokat, Y., \& Wicas, N. W. (2007). Portfolio rebalancing in theory
and practice. The Journal of Investing, 16(2), 52-59.
\item Urquhart, A. (2022). Under the hood of the Ethereum blockchain. Finance
Research Letters, 47, 102628.
\item Walpole, R. E., Myers, R. H., Myers, S. L., \& Ye, K. (1993). Probability
and statistics for engineers and scientists (Vol. 5). New York: Macmillan.
\item Wang, S., Yuan, Y., Wang, X., Li, J., Qin, R., \& Wang, F. Y. (2018,
June). An overview of smart contract: architecture, applications,
and future trends. In 2018 IEEE Intelligent Vehicles Symposium (IV)
(pp. 108-113). IEEE.
\item Wendl, M., Doan, M. H., \& Sassen, R. (2023). The environmental impact
of cryptocurrencies using proof of work and proof of stake consensus
algorithms: A systematic review. Journal of Environmental Management,
326, 116530.
\item Zamani, E., He, Y., \& Phillips, M. (2020). On the security risks
of the blockchain. Journal of Computer Information Systems, 60(6),
495-506.
\item Zarir, A. A., Oliva, G. A., Jiang, Z. M., \& Hassan, A. E. (2021).
Developing cost-effective blockchain-powered applications: A case
study of the gas usage of smart contract transactions in the ethereum
blockchain platform. ACM Transactions on Software Engineering and
Methodology (TOSEM), 30(3), 1-38.
\item Zeng, X., Hao, N., Zheng, J., \& Xu, X. (2019). A consortium blockchain
paradigm on hyperledger-based peer-to-peer lending system. China Communications,
16(8), 38-50.
\item Zheng, Z., Xie, S., Dai, H. N., Chen, W., Chen, X., Weng, J., \& Imran,
M. (2020). An overview on smart contracts: Challenges, advances and
platforms. Future Generation Computer Systems, 105, 475-491.
\item Zhou, Q., Huang, H., Zheng, Z., \& Bian, J. (2020). Solutions to scalability
of blockchain: A survey. Ieee Access, 8, 16440-16455.
\item Zou, W., Lo, D., Kochhar, P. S., Le, X. B. D., Xia, X., Feng, Y.,
... \& Xu, B. (2019). Smart contract development: Challenges and opportunities.
IEEE Transactions on Software Engineering, 47(10), 2084-2106.
\end{itemize}

\end{document}